\documentclass[12pt]{article}
\usepackage{amssymb,slashed
}
\usepackage{amsmath}
\usepackage[all]{xy}
\newcommand{\ft}[2]{{\textstyle\frac{#1}{#2}}}
\def\Re{\mathop{\rm Re}\nolimits}
\def\Im{\mathop{\rm Im}\nolimits}

\def\rme{{\mathrm e}}
\def\rmi{{\mathrm i}}
\newcommand{\ii}{\mathrm{i}}
\def\rmd{{\mathrm d}}
\newsavebox{\uuunit}
\sbox{\uuunit}
    {\setlength{\unitlength}{0.825em}
     \begin{picture}(0.6,0.7)
        \thinlines
        \put(0,0){\line(1,0){0.5}}
        \put(0.15,0){\line(0,1){0.7}}
        \put(0.35,0){\line(0,1){0.8}}
       \multiput(0.3,0.8)(-0.04,-0.02){12}{\rule{0.5pt}{0.5pt}}
     \end {picture}}
\newcommand {\unity}{\mathord{\!\usebox{\uuunit}}}
\newcommand{\abs}[1]{\lvert#1\rvert}

\csname @addtoreset\endcsname{equation}{section}
\newcommand{\SU}{\mathop{\rm SU}}
\newcommand{\SO}{\mathop{\rm SO}}
\newcommand{\U}{\mathop{\rm {}U}}

\newcommand{\Sp}{\mathop{\rm {}Sp}}

 \newcommand{\Gl}{\mathop{\rm {}G}\ell }

\newcommand{\rSU}{\mathrm{SU}}

\newif\ifpdf
\ifx\pdfoutput\undefined
  \pdffalse
  \usepackage{cite} 
  \else
  \pdfoutput=1
   \pdftrue
  \usepackage[pdftex]{hyperref}
\fi

  \def\cC{{\cal C}} \def\cD{{\cal D}}
    
   \def\cK{{\cal K}} \def\cL{{\cal L}}
 \def\cM{{\cal M}} \def\cN{{\cal N}}  \def\cP{{\cal P}}
  \def\cR{{\cal R}} \def\mR{{\mathbb{ R}}}
 \def\mZ{{\mathbb{ Z}}}
 \def\mC{{\mathbb{C}}} 
 \def\mH{{\mathbb{H}}}  

 \def\cS{{\cal S}}   \def\cV{{\cal V}}
 \def\cW{{\cal W}}   
\newcommand{\covder}{\mathfrak{D}}

\DeclareMathAlphabet\Scr{U}{rsf}{m}{n}
\begin{document}

\begin{titlepage}
\begin{center}
\baselineskip=16pt
{\LARGE Half-BPS  cosmic   string  in   $\cN=2$ supergravity \\
in the presence of a dilaton. \\
}
\vfill
{\large  Mboyo Esole,  Kepa Sousa 
  } \\
\vfill
{\small Lorentz Institute of Theoretical Physics, Leiden
University,\\ 2333 RA Leiden, The Netherlands \\ \vspace{6pt}
}
\end{center}
\vfill
\begin{center}
{\bf Abstract}
\end{center}

{\small
  
We construct new half-BPS cosmic string solutions in  $D=4$ $\cN=2$ supergravity compatible with a consistent truncation to  $\cN=1$  supergravity  where they describe $D$-term cosmic strings. 
 The constant  Fayet-Iliopoulos term  in the $\cN=1$ $D$-term   is not put  in by hand but is geometrically engineered by a gauging in the mother $\cN=2$ supergravity theory. 
    The coupling of the $\cN=2$ vector multiplets is characterized by a cubic prepotential admitting an axion-dilaton field, a common property of many  compactifications of string theory.   
     The axion-dilaton field survives the truncation  to $\cN=1$ supergravity.
     On the string configuration the BPS equations constrain the dilaton to be  an arbitrary  constant.  All the cosmic  string solutions with different values of the dilaton have the same energy per unit length but  different lenght scales.  
 }\vspace{2mm} \vfill \hrule
width 3.cm
 {\footnotesize \noindent
e-mails: \{esole,kepa\}   {at}  lorentz.leidenuniv.nl
\\}
\end{titlepage}
\addtocounter{page}{1}
 \tableofcontents{}
\newpage

\section{Introduction}

Gauge theory solitons, like magnetic monopoles and cosmic strings, are predicted by Grand Unified Theories  (GUT). Their 
formation in the early universe has nontrivial implications for the subsequent cosmological evolution \cite{VilenkinShellard,HindmarshRE}. While  the occurence of monopoles is strongly limited by  observations because they would dominate the energy density of the universe, cosmic strings are still compatible with current data. They are no longer considered as the main seed of structure in the universe in favor of inflation, but they are  predicted  by  GUT and many inflation models.  \\

In this paper we study BPS cosmic string solutions  in the context of  $D=4$ $\cN=2$ supergravity \cite{N2DtermString}. 
Supersymmetry plays an important role in todays attempts to  unify  fundamental forces beyond the Standard Model.  In a supersymmetric theory,  BPS configurations are those  that  preserve a fraction of the supersymmetries.  They constitute  interesting  probes of the high energy  regime of  the theory as they are usually protected from quantum corrections. 
As supergravity theories are supersymmetric local field theories of gravity, it is important to analyze their    (BPS) solitonic solutions. The latter  could have  non-trivial implications for the cosmology of supersymmetry theories. They could also corresponds to degrees of freedom of a more fundamental theory.\\

A particular class of BPS solutions has attracted a lot of attention recently: {\em $D$-term strings} \cite{Davis:1997bs,Morris:1997ua,Dvali:2003zh,Binetruy:2004hh, AchucarroRY, BlancoPilladoXX,UrrestillaEH,DasguptaDW, 
N2DtermString}. 
They  are Nielsen-Olesen string-like solutions of a $D=4$ Abelian-Higgs model coupled to $\cN=1$ supergravity. The corresponding $\U(1)$ gauge symmetry is spontaneously broken by the vacuum expectation value of a Higgs field.  The Higgs mechanism is due to the presence of a $D$-term potential endowed with a constant Fayet-Iliopoulos term (FI term).
$D$-term strings preserve half of the supersymmetries: they are half-BPS objects\footnote{  It is also possible to construct half-BPS cosmic strings in $\cN=1$ supergravity which are not of the Nielsen-Olesen  type. For example, the 
{\em magnetic cosmic strings} of \cite{Gutowski:2001pd} don't require any Higgs fields,  but   only a gauge field and a $D$-term endowed with a constant FI term.}
 . 
Their  tension saturate a topological bound, the {\em Bogomolnyi bound}. 
 \\

   $D$-term strings were first constructed in $\cN=1$ global supersymmetry \cite{Davis:1997bs}. Their coupling to  $\cN=1$ supergravity was first investigated in \cite{Morris:1997ua}.
    The interest in $D$-term strings increased after the authors of \cite{Dvali:2003zh}  established that they are half-BPS in $\cN=1$ supergravity and conjectured that they represent the low energy manifestation of fundamental objects in string theory called  {\em  D-strings}  \footnote{ For a review of cosmic strings in superstring theory, see \cite{Polchinski:2004ia,Davis:2005dd,Majumdar:2005qc}. 
The possibility of having  heterotic cosmic strings have been discussed  in 
 \cite{Becker:2005pv}.}. The latter are    D-branes with  one non-compact spatial  dimension\cite{Jones:2002cv,Sarangi:2002yt,Copeland:2003bj,Jackson:2004zg,Leblond:2004uc,VanProeyen:2006mf}.
 Since then several  string theory analyses have appeared in the literature that support the conjecture \cite{DasguptaDW, Halyo:2003uu,Gubser:2004qj,Lawrence:2004sm, Martucci:2006ij}. 
However they are also indications of  limitation  of the conjecture\footnote{ 
$D$-term strings are not expected to reproduce the   scattering properties of D-strings. Indeed, gauge theory  solitons have a reconnection probability $P\approx1$, while the same quantity for  D-strings has been estimated to take values in the range $10^{-1}\lesssim P \lesssim 1$ \cite{Polchinski:2004ia,Davis:2005dd}. 
The conjecture was based on the observation that $D$-term strings were the only BPS  saturated strings   available in $\cN=1$ supergravity. By now,  other  BPS strings have been obtained with different stability behaviors. For example     semilocal strings 
\cite{UrrestillaEH, DasguptaDW}  and axionic $D$-term strings \cite{BlancoPilladoXX,Achucarro:2006ef}    have a core radius that can vary in size, a  property  that  is not  generally expected for D-strings.  }. \\

As a $D$-term string requires a  constant FI term,  we are constrained to Abelian gauge theories since FI terms are only allowed in this case. FI terms are also interesting in cosmology aside from the  study of topological defects.  In the framework of $\cN=1$ supergravity they can be used to generate a positive cosmological constant, what leads to de Sitter vacua  and are crucial for $D$-term inflation \cite{Binetruy:2004hh}. \\

So far, it has not been possible to identify a mechanism in string/M-theory leading  to a constant FI term in $\cN=1$ supergravity after compactification. 
In order to identify such a mechanism, it can be useful to work in $\cN=2$ supergravity as an intermediate  step. Indeed, many compactifications of string theory  lead to $\cN=2$ supergravity in 4 space-time dimensions.
 This is an  invitation to identify the $\cN=2$ supergravity actions  which lead to  $\cN=1$  supergravity  with a  $D$-term  potential and a constant FI term. \\

 In flat space, the embedding of $D$-term potential with a constant FI term  into $\cN=2$ global supersymmetry is well understood \cite{Kallosh:2001tm}.   $\cN=2$ global supersymmetry admits a scalar potential that depends on a triplet of {\em moment maps} that generalize the $D$-term of $\cN=1$ susy. In flat space, the triplet of moment maps  admits a  triplet of FI terms that generalize those  of $\cN=1$ susy. \\

 The situation changes drastically when we consider the coupling to gravity. Indeed,   $\cN=2$ supersymmetry forbids constant FI terms in the presence of hypermultiplets \footnote{ As we shall review later on, this is due to a topological obstruction coming from the geometry of hypermultiplets in $\cN=2$ supergravity \cite{Kallosh:2001tm,Andrianopoli:1997cm}}.  
  At first sight this seems to exclude a description of $\cN=1$ 
 $D$-term potential with constant   FI terms from $\cN=2$ supergravity  in presence of   hypermultiplets. However,  FI terms in $\cN=1$ supergravity do not need to originate  from FI terms in $\cN=2$ supergravity.  In particular it is possible that a  scalar potential in a $\cN=2$ action, after truncating consistently part of the fields, will correspond to an $\cN=1$ potential with a $D$-term and a constant FI  term\footnote{  
 An action   is said to be consistently truncated to a reduced  action, when any solution of the equations of motion of the reduced action  is also a solution of the full action.}.\\

The first  known example of a half-BPS cosmic string in $\cN=2$ supergravity was  constructed in   \cite{N2DtermString}. It involves   the minimal matter content needed to obtain a half-BPS cosmic string solution in $\cN=2$ supergravity action: one hypermultiplet and one vector multiplet\footnote{ The hypermultiplet had to be included to provide the scalar acting as a Higgs field since, for Abelian gauging, supersymmetry forbids the scalars of  vector multiplets to be charged under gauge transformations.}. 
The construction of \cite{N2DtermString} can be seen as a consistent embedding of an $\cN=1$ half-BPS cosmic string into $\cN=2$ supergravity while preserving the half-BPS nature of the object.  \\

It is useful to recall some  aspects of $\cN=2$ supergravity in order to understand the construction of \cite{N2DtermString} and the  purpose  of the  present paper. \\

The scalar fields in $\cN=2$ supergravity can be seen as coordinates of a   {\em scalar manifold} 
\begin{equation}
\mathrm{M}=\mathrm{M}_V \otimes \mathrm{M}_H,\nonumber
\end{equation}
 where $\mathrm{M}_V$ and $\mathrm{M}_H$ correspond respectively to the scalar fields of vector multiplets and hypermultiplets. 
The constraints coming from supersymmetry can then be understood as geometric conditions on the scalar manifold $M$.  $\cN=2$ supergravity requires  $\mathrm{M}_V$ to be a Special-K\"ahler and 
$\mathrm{M}_H$ to be of Quaternionic-K\"ahler type.  \\

The kinetic terms of the scalar fields are defined by a sigma model with target space the scalar manifold $\mathrm{M}$.  In order to respect supersymmetry, all internal symmetries of the theory have to preserve the kinetic terms of the scalar fields and therefore  correspond to  isometries of the scalar manifold. 
When a group of these  internal symmetries is promoted to be {\em local  group} by letting the transformation parameters of the symmetries to depend on spacetime, it is said that the symmetry group has been {\em gauged}. A gauging introduces new couplings between scalar and gauge fields coming from the usual {\em minimal coupling} which consists of replacement of derivatives by  covariant derivatives. When a supergravity theory is gauged, more couplings have to be included in the action to preserve supersymmetry. In particular, the  supersymmetry transformations of the fermions are modified by the so-called {\em fermionic shifts} that also   play the role of  mass matrices for the fermions.  In $\cN=2$ supergravity, a gauging also requires the introduction of  a scalar potential     quadratic in the fermionic shifts. \\

In $\cN=1$ supergravity, the $D$-term corresponds to  the part of the scalar potential generated by a gauging   whereas the $F$-term is due to a superpotential. The $D$-terms are  fermionic  shifts for the gaugini. The superpotential gives masses to the gravitino and the chiral fermions. 
In $\cN=2$ supergravity (like in all extended supergravity theories), gauging is the only way to generate a scalar potential\footnote{When hypermultiplets are not present, it is possible to include a triplet of $\cN=2$ constant Fayet-Iliopoulos term without any gauging. }. \\

In the present paper we shall only be concerned with Abelian gauging.
When considering Abelian gauging, only the isometries of the quaternionic-K\"ahler manifold $\mathrm{M}_H$ are relevant as the scalar fields of the special manifold $\mathrm{M}_V$, being part of vector multiplets, are neutral under Abelian symmetries.  
A  gauging of  isometries of the quaternionic-K\"ahler manifold $\mathrm{M}_H$ contributes to the  fermionic shifts through the Killing vectors representing the isometries  and a triplet of so-called  {\em moment-maps}. \\

In  \cite{N2DtermString}, after choosing the scalar manifold $\mathrm{M}$, a specific compact $\U(1)$ symmetry of the quaternionic manifold was gauged to generate a scalar potential.
The ansatz for the cosmic string solution used only a subset of the fields in such a way that the solution  was also valid for an   $\cN=1$ supergravity model coupled to one vector fields and one chiral multiplet. In other words, the cosmic string ansatz is   compatible with a consistent truncation of the theory to $\cN=1$ supergravity.   However, supersymmetry is not spontaneously broken to $\cN=1$ supergravity and the full $\cN=2$ supersymmetry is preserved in the vacuum at spatial infinity. 
From a $\cN=2$ point of view, the use of an ansatz related to a  consistent truncation to $\cN=1$ supergravity in \cite{N2DtermString}   is not required but is useful  to   simplify the calculation, as $\cN=2$ BPS equations are in general much more difficult to solve than those of  $\cN=1$ theories. 
Alternatively, one can consider that the construction of \cite{N2DtermString} is a consistent embedding of a cosmic string solutions of an  $\cN=1$ supergravity model into an $\cN=2$ supergravity model while preserving the BPS property of the solution.
Consistent truncations of $\cN=2$ supergravity are studied systematically in \cite{Andrianopoli:2001zh,Andrianopoli:2001gm,D'Auria:2005yg}.  \\

  The bosonic part of the reduced $\cN=1$ action  of \cite{N2DtermString}  is 
\begin{eqnarray}
e^{-1}\cL= \frac{1}{2} R  -\frac{1}{4}F^{\mu\nu}F_{\mu\nu} 
-\frac{1}
{2(\Im \Phi)^2} \nabla  _\mu \Phi\nabla  ^\mu \bar \Phi
- 2 g^2\left[   \frac{\abs{\Phi}^2+1}{\Im \Phi}-\xi  \right]^2,\nonumber
\end{eqnarray}
where
$F_{\mu\nu}=\partial_\mu W_\nu-\partial_\nu W_\mu ,\quad  \nabla _\mu\Phi =\partial _\mu \Phi-2gW_\mu \left( \Phi ^2+1\right)$.
There is no superpotential, the kinetic tem of the vector fields has a trivial metric and the $D$-term is $D=\frac{\abs{\Phi}^2+1}{\Im \Phi}-\xi$, where $\xi$ is the FI term.\\

The main purpose of the present  paper is to enlarge the type of $\cN=2$ supergravity theories that can generate constant FI term in $\cN=1$ supergravity. The generalization to other quaternionic manifolds that are symmetric spaces is straightforward. However, the special geometry was a very particular case. 
In $\cN=2$ supergravity, special geometry determines the couplings of the vector fields of the theory.  The special geometry used in \cite{N2DtermString} corresponds to  the so-called {\em minimal special geometry}. \\

In special geometry, the scalar manifold is a K\"ahler-Hodge manifold\footnote{A K\"ahler-Hodge manifold is a K\"ahler manifold with a K\"ahler form defining an integer cohomology. In $\cN=1$ supergravity, the scalar manifold can be any   K\"ahler-Hodge manifold.  For a   special K\"ahler-manifold, the K\"ahler two-form  has an even integer cohomology.}. However,   the Kahler potential $\cK$ is not the fundamental object. It is computed in terms of a so-called {\em symplectic section }   $(Z^I, F_I)$  as 
\begin{equation}
\cK= -\log \left[{-\ii\left( Z^I  \bar F_I-  F_I \bar Z^I \right)}\right].\nonumber
\end{equation}

where $I=0,\ldots, n_V$ and $n_V$ is the number of vector multiplets. The section  depends on the scalar fields of the vector multiplets. \\

The symplectic section is subject to symplectic transformations. The latter are  not symmetries  of the Lagrangian but of the equations of motion and Bianchi identities of the vector fields.
The K\"ahler potential is also a symplectic invariant and therefore the geometry of the special manifold is invariant under symplectic  rotations.  However, the scalar potential and the metric of the kinetic term of the vector fields are not invariant under symplectic rotations.

A symplectic section is said to admit a {\em prepotential} when it is possible to define an holomorphic function $F(Z)$ depending only on the first half of the symplectic section such that 
\begin{equation}
F_I=\frac{\partial}{\partial Z^I} F.  \nonumber
\end{equation}
$F(Z^I)$ is called the {\em prepotential} and is required to be homogeneous of order two (that is  $F(\lambda Z)=\lambda^2 F(Z)$, for any complex $\lambda$).  It  is always possible to rotate a given symplectic section into one that admits a prepotential \cite{Craps:1997gp}. It follows that prepotentials provide a simple way to classify families of  special geometries. Such a classification is enough to discuss aspects of special geometry that are invariant under symplectic rotations like for example the geometry of the special manifold. However, more generally,   the prepotential is not enough to distinguish two physically different supergravity theories once the theory is gauged. For example, spontaneous supersymmetry breaking to $\cN=1$ supersymmetry is only possible when the symplectic section does not admit a prepotential \cite{Ferrara:1995gu}. Moreover, as the scalar potential is not a symplectic invariant in $\cN=2$ supergravity, the stability of vacua can be modified by a symplectic rotation as illustrated in the study of de Sitter vacua in \cite{Fre:2002pd}. \\

Minimal special geometry corresponds to a quadratic prepotential $F$:
\begin{equation}
F(Z)=-\frac{\ii}{2}(Z^0 Z^0-Z^a Z^a), \quad a=1,\ldots, n_V.  \nonumber
\end{equation}
This is the so-called {\em minimal prepotential} and the  corresponding special manifold  is $\frac{\SU(1,n_V)}{\U(n_V)}$.

 Minimal special geometry has many virtues which makes it an interesting  candidate  to  construct  cosmic string solutions. In particular,  all the scalar fields of the special manifold can be consistently truncated on the string configuration, in such a way that the string solution is based on the minimal amount of matter field required:  one vector field and one scalar field. The metric of the kinetic terms of the vector fields is then trivial.\\

To enlarge the family of supergravity models related to $\cN=1$ supergravity with a $D$-term, and admitting a constant FI term, we shall study the  mechanism described in \cite{N2DtermString}  with a special geometry  that is  related to a {\em cubic prepotential}:
\begin{equation}
F=\ii d_{IJK}\frac{Z^I Z^J Z^K}{Z^0}, \quad I , J, K= 1,\ldots, n_V, \nonumber
\end{equation} 
where  $ d_{IJK}$ are  real constant coefficients  symmetric in their three indices. \\

 Realizing the construction of \cite{N2DtermString} with a special geometry related to a cubic prepotential is a natural direction for generalization in view of all the interesting models that require this type of special geometry.
 Cubic prepotentials are classified in \cite{deWit:1991nm}. They   characterize the $\cN=2$, $D=4$ supergravity theories coming from $\cN=2$, $D=5$ supergravity theories. They are also the special geometry of  many compactifications of string theory  like for example type $II$ string theories compactified on  Calabi-Yau three-folds \cite{Andrianopoli:1997cm,Ceresole:1995jg,Craps:1997gp} and the Heterotic string compactification on $K3\times T^2$  \cite{Andrianopoli:1997cm,Ceresole:1995jg,Strominger:1990pd}  and   the $D3-D7$ model (type $IIB$ on $K3\times T^2/\mZ_2$ in presence of open string moduli) \cite{Andrianopoli:2003jf,Angelantonj:2003zx,Hsu:2004hi}. As we shall see, the change of the symplectic section will imply important differences in the qualitative  behaviour of the resulting cosmic string solution. Our choice of special geometry will contain an axion-dilaton field  $S=a-\ii \rme^\rho$ parametrizing the K\"ahler manifold $\frac{\SU(1,1)}{\U(1)}=\frac{\SO(2,1)}{\SO(2)}=\frac{\mathrm{SL}(2,\mR)}{\SO(2)}$  which corresponds to the complex half-plane.\\

We will consider the special geometry based on the coset space   
\begin{equation}
\mathrm{M}_V = ST[2,n] \equiv \frac{\SU(1,1)}{\U(1)}\times \frac{\SO(2,n)}{\SO(2)\times \SO(n)}, \nonumber
\end{equation} 
with the  so-called {\em Calabi-Visentini symplectic section}   \cite{Andrianopoli:1997cm,Ceresole:1995jg}  well-known from different  compactifications of string theory    \cite{Andrianopoli:2003jf,Angelantonj:2003zx}. \\

On the quaternionic side,  the analysis can be done   with any  normal quaternionic manifold. To avoid  complications that are not essential to the construction, we  shall consider the  quaternionic manifold of (quaternionic)   dimension one  $\mathrm{M}_H = \frac{\SO(4,1)}{\SO(4)}$.  The model is then simple enough to be analyzed in   detail and our results can be compared with those of \cite{N2DtermString}.\\

We perform the  same  compact Abelian gauging as in \cite{N2DtermString} and use a consistent truncation ansatz to obtain cosmic string solutions. The reduced theory is an $\cN=1$ supergravity coupled to a vector multiplet and two chiral multiplets corresponding to the axion-dilaton and Higgs fields.\\

We shall see that with the choice of the Calabi-Visentini symplectic section, the  $\cN=2$ scalar potential is bounded from below as long as we gauge a  vector of  negative signature with respect to the metric $\eta_{I  J }$ used to define the Calabi-Visentini basis.  
In the case of minimal special geometry it was possible to truncate all  the scalar fields of the vector multiplets   on the string configuration. In the present case, there is a neutral   axion-dilaton field, $S=a-\ii\rme^\rho$,
parametrizing the sub-manifold $\frac{\SU(1,1)}{\U(1)} \subset \mathrm{M}_V$,  which survives on the string configuration   together with the vector field and the Higgs field.\\

The bosonic part of the truncated $\cN=1$ action, as defined above, is: 
\begin{eqnarray}
e^{-1}\cL&=& \frac{1}{2} R 
 + \frac{1}{4}(\Im S) F^{\mu\nu}F_{\mu\nu} 
+\frac{e^{-1}}{8}(\Re S) \varepsilon^{\mu\nu\rho\sigma}F_{\mu\nu}F_{\rho\sigma}\nonumber\\
& & 
-\frac{1}
{4(\Im S)^2} \partial _\mu S \partial  ^\mu \bar S-\frac{1}
{2(\Im \Phi)^2} \nabla  _\mu \Phi\nabla  ^\mu \bar \Phi \nonumber \\
&&+ \frac{2g^2}{\Im S}\left[       \frac{\abs{\Phi}^2+1}{\Im \Phi}-\xi  \right]^2, \nonumber
\end{eqnarray}
where
$F_{\mu\nu}=\partial_\mu W_\nu-\partial_\nu W_\mu ,\quad  \nabla _\mu\Phi =\partial _\mu \Phi-2gW_\mu \left( \Phi ^2+1\right)$.   The holomorphic function that defines the kinetic term of the vector field is $f=\ii S$ where  $ S=a-\ii \rme^{\rho}$ is the axion-dilaton field.  
The K\"ahler-Hodge manifold is 
\begin{equation}
\left({\frac{\SO(2,2)}{\SO(2)\times \SO(2)}}\right)_{S,\Phi}=\left({\frac{\SU(1,1)}{\U(1)}}\right)_{S}\times \left({\frac{\SU(1,1)}{\U(1)}}\right)_{\Phi}, \nonumber
\end{equation}
Each of the two complex scalar fields ($S$ and $\Phi$) parametrize  a $\frac{\SU(1,1)}{\U(1)}$ factor. The K\"ahler-potential is $\cK=-\log\left[{\ii(S-\bar S)}\right]-2 \log\left[{-\ii(\Phi-\bar\Phi)}\right]$.
 There is no superpotential and the   $D$-term is $D=\frac{\abs{\Phi}^2+1}{\Im \Phi}-\xi$, where $\xi$ is the FI term.The gauge symmetry is $\delta \Phi=2 g (\Phi^2+1)$.
\\

 We shall show that it is possible to obtain half-BPS cosmic string solutions which solve the full $\cN=2$ equations  of motion. These string solutions are defined in the absolute minimum of the potential, which  is of Minkowksi type and preserves the full $\cN=2$ supersymmetry. \\
 
 The  $\cN=2$ BPS equations imply  that the  axion-dilaton is an arbitrary constant. 
For a fixed winding number, we shall show that all the half-BPS solutions with different values of the dilaton have the same energy per unit length.  \\

Solutions with different value of the dilaton can be distinguishable by their characteristic length scales.  
Indeed, the masses of the fields that define the string solution, $m_W$ for the vector and $m_\Phi$ for the scalar, depend explicitly on the dilaton field:
\begin{equation}
\frac{1}{m^2_W} \equiv l_W^2 \propto  -\Im S =\rme^\rho , \quad  \frac{1}{m^2_\Phi} \equiv l_\Phi^2 \propto -\Im S= \rme^\rho. \nonumber
\end{equation}
As a consequence we obtain a one parameter family of string solutions, degenerate in energy but with varying core radius.\\

The structure of the paper is the following.
In section 2 we review the mathematical tools needed for $\cN=2$ supergravity: special and quaternionic geometry, the gauging of isometries and the $\cN=2$ supersymmetry transformations. In section 3, after discussing consistent reduction of supersymmetry, we explain how the constant FI term is obtained in the reduced $\cN=1$ supergravity model.
In section 4 our model is presented. We give the choice of special and quaternionic geometry together with the gauging, which determines the scalar potential.
In section 5 we study the vacua of the potential and the
$\cN=1$ reduced theory is presented. Next, we construct the string ansatz and we solve the BPS equations. At the end of this section some of the properties of the string are discussed.
Finally we review our results in section 6.

\section{Review of $\cN=2$ Supergravity}

\subsection{Overview}

 Supersymmetry is a symmetry that transforms bosons and fermions into each other. It follows that supersymmetry transformations admit fermionic generators.   In $D=4$ dimensional spacetime with Minkowski signature, the supersymmetry generators are Majorana spinors.  A Majorana spinor admits four  degrees of freedom and can be decomposed into  two chiral spinors of opposite chirality. 
In four dimensional spacetime,  a {\em $\cN$ extended supersymmetric} theory admits $4\cN$ supersymmetric generators organized into $\cN$ Majorana spinors.
      $\cN=2$ supersymmetry requires two Majorana spinors that represent a total of 8 independent supersymmetry generators. We shall work with the corresponding four chiral spinors that we denote $(\epsilon^i, \epsilon_i)$. The index $i=1,2$  labels the original Majorana spinors and the position of that index represents the chirality. $\epsilon^i$ is a left-handed spinors while  $\epsilon_i$ is right-handed:
 \begin{equation}
\epsilon^i =\ft12(1+\gamma_5)\epsilon^i,\quad  
\epsilon_i =\ft12(1-\gamma_5)\epsilon_i.
\end{equation}
 We follow the notation and conventions of \cite{LectParis,VanProeyen:1999ni}.
 Charge conjugation relates the two chiral projections of a given Majorana spinor. 
 We shall use the same convention for other chiral spinors.\\

  The automorphism group  of the 
 supersymmetry algebra, the so-called  {\em $\mathrm{R}$-symmetry group},
  rotates the supersymmetry generators among themselves while preserving the supersymmetry algebra  and commuting with the Lorentz transformations. The ${\mathrm{R}}$-symmetry group  plays an important  role in determining the  structure of supersymmetric theories.  In four spacetime dimensions with the usual Minkowski signature, the $\mathrm{R}$-symmetry group is
  $H_{\mathrm{R}}=\U(\cN)$ where $\cN$ is the number of supersymmetries. In particular, we have 
 $H_{\mathrm{R}}=\U(1)$  for $
 \cN=1$ supergravity and a $H_{\mathrm{R}}=\U(2)=\U(1)\otimes \SU(2)$ for $\cN=2$ supergravity. 
 Under the $\SU(2)$ part of the $\mathrm{R}$-symmetry, the supersymmetry generators $\epsilon^i$ transform as a doublet. The $\U(1)$ part of the $\mathrm{R}$-symmetry acts on the generators by a change of   phase. \\
 
 The matter content of a supersymmetric theory is organized into  irreducible representations of the 
super-Poincar\'e group called {\em (super)multiplets}. Inside each multiplet, the fields are
 arranged into representations of the $\mathrm{R}$-symmetry group. \\

 Supergravity theories are field theories endowed with local supersymmetry, that is, the supersymmetry generators are  spacetime dependent.
 In supergravity, the  algebra of supersymmetry transformations contains the Poincar\'e group and therefore supergravity theories always contain gravity. The latter is represented by a graviton multiplet which contains the vielbein of the spacetime metric and its fermionic partners, the gravitini. 
 A $\cN$ extended supergravity theory has $\cN$ gravitini in the graviton multiplet.\\
 
 In $\cN=2$ supergravity, we shall consider  three type of multiplets: 

\begin{itemize}
\item the graviton multiplet : it contains the vielbein of the spacetime metric, two gravitini $\psi^i_\mu$ and one graviphoton $W_\mu^0$. The label $i=1,2$ is associated to the $\SU(2)$ $\mathrm{R}$-symmetry transformations. 

\item the vector multiplet: it contains one complex scalar $z^\alpha$, two gaugini $\lambda ^\alpha_ i$  and one gauge field $W_\mu^\alpha$.  Here $\alpha=1, \ldots , n_V$ labels $n_V$ different vector multiplets.

\item the hypermultiplet: it contains four real scalars $q^X$  and two hyperini,  $\zeta ^A$, where the labels are  $X=1, \ldots , 4 n_H$ and $A=1, \ldots , 2 n_H$ for  $n_H$  hypermultiplets.
\end{itemize}

\begin{table}[h]
\begin{center}
\begin{tabular}{| l | c | c | c |}
\hline 
& &  & \\
& vielbein & $e^a_\mu$ & \\
& &  & \\
Gravity multiplet &  gravitini & $\psi^i_\mu,\psi_{i\mu}$&  $\begin{matrix}  i=1,2 \\ \mu,a=0,\cdots,3\end{matrix}$ \\
& &  & \\
& graviphoton & ``$W^0_\mu$''   & \\
& & & \\
\hline 
\hline
&  & & \\
&  gauge fields & $W_\mu^\alpha$& \\
&  & & \\
Vector multiplet &  gaugini & $\lambda^\alpha_i,\lambda_\alpha^i$ & $ \alpha=1,\ldots, n_V$ \\
&  & & \\
& scalars & $z^\alpha$ & \\
& &  & \\
\hline
\hline
 & & & \\
 & hyperscalars & $q^X$ & \\
  Hypermultiplets & & &  $\begin{matrix}X=1,\ldots, 4n_H\\ A=1,\ldots, 2 n_H\end{matrix}$\\
  & hyperini &$\zeta^A,\zeta_A$  &  \\
& & & \\
  \hline
\end{tabular}
\caption{Field content of $\cN=2$ supergravity coupled to $n_V$ vector multiplets and $n_H$ hypermultiplets. The physical graviphoton  is not necessarily $W^0_\mu$ but whatever field
appears in the supersymmetric transformation of the gravitini through  its field strength $T_{\mu\nu}$. The latter is   a linear combination of field strengths of all the gauge fields  $W_\mu^I$ present in the theory ($I=0,\ldots, n_V$)  with coefficients that depend on the scalar fields $z^\alpha$ of vector multiplets. The couplings of  all the gauge fields $W^I_\mu$ and the scalar fields $z^\alpha$ is controlled by {\em special geometry}. }
\end{center}
\end{table}

\subsubsection{Supersymmetry and geometry}

The scalar fields present in supersymmetric multiplets  can be seen locally as coordinates of  a manifold, (the {\em scalar manifold} $\mathrm{M}$), whose geometry is restricted by supersymmetry. The latter splits into a direct product of scalar manifolds corresponding to  different types of multiplets present in the theory.\\

When there is an action, the  kinetic terms of the scalar fields $\phi^r$ define a sigma model with target space the scalar  manifold $\mathrm{M}$
\begin{equation}
\cL_{\phi,\, \text{kinetic}}= -\ft{1}{2}g_{rs}(\phi)
\partial_\mu \phi^r\partial^\mu \phi^s.
\end{equation}
As we have said before,  the scalar fields $\phi^r$ can be seen as  local coordinates of  the scalar manifold 
$\mathrm{M}$. From this point of view, $g_{rs}(\phi)$ is interpreted as a metric defined on $\mathrm{M}$. Thus the scalar manifold $\mathrm{M}$ has the structure of a  Riemannian space. \\ 
 
The supersymmetry transformations will involve  a vielbein on the scalar manifold $\mathrm{M}$. 
The reality conditions due to the type of spinors that are used, the $\mathrm{R}$-symmetry and the closure of the supersymmetry algebra will impose restrictions on the vielbein. These restrictions will be understood as geometric conditions on the scalar manifold. \\

In $\cN=2$ supergravity coupled to $n_V$ vector multiplets and $n_H$ hypermultiplets, the scalar manifold is a direct product 
\begin{equation}
\mathrm{M}=\mathrm{M}_V \otimes \mathrm{M}_H,
\end{equation}
where  $\mathrm{M}_V$ and  $\mathrm{M}_H$ are respectively the scalar manifold of vector and hypermultiplets. The restrictions coming from supersymmetry impose that $\mathrm{M}_V$ is a so-called {\em special manifold}, whereas $\mathrm{M}_H$ is a {\em quaternionic manifold}. \\

For convenience, the $\cN=2$ supersymmetry transformations will be reviewed later on, after the geometry of the scalar manifold and the gauging of isometries have been discussed.

\subsection{Vector multiplets and special geometry}
We consider $\cN=2$ supergravity coupled to  $n_V$ vector multiplets \cite{deWit:1984pk,deWit:1984px,Strominger:1990pd}. For a modern review, see  \cite{Andrianopoli:1997cm,Craps:1997gp,deWit:1995jd}.  \\

Since the gravity multiplet contains a vector field, the {\em graviphoton},  this theory admits $n_V+1$ vector fields   $W^I_\mu$ where $I=0,\ldots, n_V$. 
The $n_V$ vector multiplets contain as well $n_V$ complex scalar fields $z^\alpha$, $\alpha=1,\ldots, n_V$ parametrizing a  K\"ahler manifold. 

The kinetic terms of the scalar  and vector fields are :
\begin{eqnarray}
  e^{-1}{\cal L}_{vector}&=&
  \frac{1}{4}\Im (\cN)_{IJ}F^{I\mu\nu}F^J_{\mu\nu}-\frac{1}{8} e^{-1}\Re(\cN)_{IJ}\epsilon^{\mu\nu\rho\sigma}F^I_{\mu\nu}F^J_{\rho\sigma}
-g_{\alpha \bar \beta }\partial_\mu z^\alpha \partial ^\mu \bar z^{\bar \beta }
\nonumber,\\
&=&  
  \ft12\Im \left( {\cal N}_{IJ}F_{\mu \nu }^{+I}F^{+\mu \nu J}\right)
  -g_{\alpha \bar \beta }\partial_\mu z^\alpha \partial ^\mu \bar z^{\bar \beta },
 \label{Lvector}
\end{eqnarray}
where $ F_{\mu \nu }^{\pm I}$ is the self-dual combination \footnote{In our convention the Levi-Civita tensor satisfies $\varepsilon_{0123}=1$.}:
\begin{align}
  F_{\mu \nu }^{\pm I} =\ft12\left( F^I_{\mu \nu }\mp \ft12 \rmi e\varepsilon_{\mu \nu \rho \sigma
  }F^{I\rho\sigma } \right) ,\qquad F^I_{\mu \nu }=\partial _\mu W_\nu ^I-\partial _\nu W_\mu ^I.
 \label{defF+-}
\end{align}

The matrix $\cN_{IJ}$ is a function of the scalar fields $z^\alpha$. Its real and imaginary parts generalize   the  inverse of the coupling constant and the $\theta$-parameter familiar from the Abelian gauge theory: 
\begin{equation}
e^{-1}\cL=-\frac{1}{4g^2}F^{\mu\nu}F_{\mu\nu}+e^{-1}\theta\epsilon^{\mu\nu\rho\sigma}F_{\mu\nu}F_{\rho\sigma}.
\end{equation}
 The couplings of vector multiplets to $\cN=2$ supergravity are elegantly expressed by 
 {\em  special geometry}. The latter relies heavily  on the existence of {\em duality transformations} for  vector fields  in supersymmetric theories. 
 Duality transformations generalize the electric-magnetic duality of Maxwell's equations without sources in the sense that they   are  linear transformations ${\cal S}\in \Gl(2n_V+2, \mR)$:
  \begin{equation}
\begin{pmatrix}
\tilde F^+\\
\tilde G_+
\end{pmatrix}=
{\cal S}
\begin{pmatrix}
 F^+\\
G_+
\end{pmatrix},
\end{equation}
of the vector field strengths {\bf $F^{+ I}$} and their magnetic duals
\begin{equation}
G^{\mu\nu}_{+I}= 2\ii \frac{\partial \cL}{\partial F^{+I}_{\mu\nu}}=\cN_{IJ}F^{+J\mu\nu}, 
\label{duality}
\end{equation}
 that  preserve the Bianchi identities and equations of motion for vector fields:
   \begin{align}
\partial^\mu \Im F^{+I}_{\mu\nu} & =0, \quad \text{Bianchi Identity},\nonumber \\
\partial_\mu \Im G^{\mu\nu}_{+I} &=0, \quad  \text{Equations of motion}.
\end{align}

Duality transformations should preserve the relation  \eqref{duality}. 
This requires that the coupling matrix $\cN_{IJ}$ undergoes the following fractional transformation:  
\begin{equation}
\cN \Longrightarrow (C+D\cN)(A+B\cN)^{-1},\label{Ntransformation}
\end{equation}
under a duality transformation given by a general invertible linear operator 
$\cS=
\begin{pmatrix}
A & B \\
C &  D
\end{pmatrix}\in \Gl(2n_V+2,\mR)$  where  $A,B,C,D$ are $(n_V+1)\times (n_V+1)$ real matrices.

If equations \eqref{duality} are derived from a Lagrangian $\cL$, the matrix $\cN_{IJ}$ should be symmetric.  Asking the symmetry of $\cN_{IJ}$ to be preserved under a fractional transformation  \eqref{Ntransformation} restricts the   duality transformations to be given by  symplectic matrices ${\cal S}$ :
\begin{equation}
\cS
\in \Sp(2n_V+2,\mR).
\end{equation}

As  the coupling matrix  $\cN_{IJ}$ transforms under duality transformations, the scalar fields  $z^\alpha$ should also transform in a specific way. 
  The action of the  duality transformations on the scalar fields is much more transparent once we introduce a {\em symplectic section} which depends on the scalar fields of vector multiplets and transforms linearly under symplectic rotations.    Special geometry can be  completely defined in terms of this symplectic section.

The symplectic section is given by:  
\begin{equation}
v=\begin{pmatrix}
Z^I \\
F_I
\end{pmatrix}, \quad I=0,\cdots n_V
\end{equation}
and is endowed with a  symplectic scalar product  
\begin{equation}
 \langle v|\bar v\rangle = -
 v^T  \begin{pmatrix}
\mathbf{0}_n & -\unity_n\\
\unity_n & \mathbf{0}_n 
\end{pmatrix} \bar v.
\end{equation}
Here $Z^I$ and $F_I$ are functions of the coordinates $z^\alpha$  ($\alpha=1,\ldots, n_V$) of the scalar fields of the vector multiplets.
Recall that the index $I$ runs from $0$ to $n_V$  where $n_V$ is the number of vector multiplets whereas $\alpha=1,\ldots,n_V$  because the graviphoton that appears in the graviton multiplet is not related to any scalar fields. This is compensated by the freedom to re-scale the symplectic section : the symplectic section is a projective section. \\

Symplectic rotations act linearly on the symplectic section as: 
\begin{equation}
\begin{pmatrix}
Z^I \\
F_I 
\end{pmatrix}\Longrightarrow 
\cS
\begin{pmatrix}
Z^I \\
F_I 
\end{pmatrix}.
\end{equation}

We see that the upper part $Z^I$ and the lower part  $F_I$ of the symplectic section transform respectively as the field strengths $F^{\pm I}_{\mu\nu}$ and their magnetic duals $G^{\pm I}_{\mu\nu}$.
This can be understood from the following remark:  $Z^I$ and $F_I$ are the  fermi-fermi components of the superspace generalization of electric and magnetic field strenghts $\hat F^I_{\mu\nu}$ and $\hat G_{I\mu\nu}$:
\begin{align}
\hat F^I_{\mu\nu  } &=F^I_{\mu\nu}+\bar Z^I \bar\psi^i_{\mu}\psi^j_{\nu}\varepsilon_{ji}+  Z^I \varepsilon^{ji}\bar\psi_{i\mu}\psi_{j\nu},\nonumber \\
\hat G_{I\mu\nu  } &=G_{I\mu\nu}+
\bar F_I \bar\psi^i_{\mu}\psi^j_{\nu}\varepsilon_{ji}+  F_I \varepsilon^{ji}\bar\psi_{i\mu}\psi_{j\nu}.
\end{align}

A special manifold is a K\"ahler manifold in which  the  K\"ahler potential is not a fundamental quantity but is  given by the following symplectic invariant expression: 
\begin{equation}
\cK=-\log \left({-\ii \langle v|\bar v \rangle}\right) = -\log \left[{-\ii\left( Z^I \bar F_I- F_I\bar Z^I \right)}\right].
\end{equation}
Here we see that the freedom to  re-scale  the symplectic section  corresponds to a  K\"ahler transformations. Under a K\"ahler transformation generated by an holomorphic function $f$,  fermions are subjects to  chiral rotations:
\begin{equation}
\lambda_i^\alpha \Longrightarrow \rme^{\frac{\ii}{2} \Im f} \lambda_i^\alpha,\quad 
\lambda^i_\alpha \Longrightarrow \rme^{-\frac{\ii}{2} \Im f} \lambda^i_\alpha.
\end{equation}
 In order for the fermions to be globally defined all over the manifold in presence of such chiral rotations, the K\"ahler form 
\begin{equation}
 K=\frac{1}{2 \pi}g_{\alpha\bar \beta}z^\alpha\wedge z^{\bar \beta}=\frac{1}{2 \pi}\partial_\alpha\partial_{\bar \beta}\cK \rmd z^\alpha\wedge z^{\bar \beta},
 \end{equation}
 should define an even coholomology, that is  
 \begin{equation}
 c_1=\frac{1}{2}[K]\in \mZ .
 \end{equation}
By definition,this condition means that the special manifold is a so-called  {\em K\"ahler-Hodge manifold}. This type of geometry  first appeared in physics as scalar manifolds for chiral multiplets in $\cN=1$ supergravity \cite{Bagger}. \\  

   In special geometry, the kinetic terms of both scalar and vector fields are computed from the symplectic section as follow 
\begin{equation}
  g_{\alpha \bar \beta }=\partial _\alpha \partial _{\bar \beta } {\cal
  K}(z,\bar z)=\rmi \langle {\cal D}_\alpha v|{\cal D}_{\bar \beta }\bar
  v\rangle,\qquad
{\cal N}_{IJ}\equiv \left( \begin{array}{cc}F_I &\bar {\cal
D}_{\bar\alpha}\bar F_I
\end{array}\right)  \left( \begin{array}{cc}Z^J &\bar
{\cal D}_{\bar\alpha}\bar Z^J
\end{array}\right)^{-1}. 
\end{equation}
Here the covariant derivatives are defined by
\begin{equation}
  {\cal D}_\alpha v=\partial _\alpha v+(\partial _\alpha {\cal K})v,\qquad
{\cal D}_{\bar \alpha} \bar v=\partial_{\bar \alpha}\bar
v+(\partial_{\bar \alpha} {\cal K})\bar v.
\end{equation}

In contrast to the metric $\cN_{IJ}$ of the vector fields,  the metric of the scalar fields is a symplectic invariant quantity.  The Riemann tensor of a special manifold is determined by a symplectic invariant and a holomorphic totally symmetric tri-tensor  $\cW_{\alpha\beta\gamma}$: 
\begin{equation}
R^\alpha{}_{\beta \gamma}{}^\delta=
\delta^\alpha_\beta \delta^\delta_\gamma +\delta^\alpha_\gamma \delta^\delta_\beta 
-\rme^{2\cK}\cW_{\beta \gamma \epsilon} \bar\cW^{\epsilon\alpha \delta},
\end{equation}
with:
\begin{equation}
\cW_{\alpha\beta\gamma}=\rme^{-\cK}\langle{\cD_\alpha U_\beta | U_\gamma} \rangle, \quad \text{where } U_\alpha = \cD_\alpha V,\quad  V=\rme^{\frac{\cK}{2}}v.
\end{equation}

When the theory is gauged, the electric-magnetic duality is explicitly broken by the introduction  of electric charges. In particular, the scalar potential generated by the gauging is not  symplectic invariant.

\subsubsection{Special geometry and prepotentials}

A special geometry is said {\em to admit a prepotential} when  the lower component of the symplectic section (the variable $F_I$) can be expressed as derivative of a scalar function $F(Z)$ depending only on the upper part of the symplectic section ( $Z^I$) :
\begin{equation}
F_I = \frac{\partial}{\partial Z^I} F(Z).
\end{equation}
$F(Z)$ is restricted to be an homogeneous function of second degree in the $Z^I$ fields and is called the {\em prepotential}.\\
 
Although prepotentials are not necessary to define special geometry, they provide a handy way to classify  special manifolds as  any symplectic section can be rotated to a section admitting a prepotential \cite{Craps:1997gp}. If one is interested only in symplectic invariant quantities, working only with prepotential is not a restriction. This is a practical approach to the classification of special geometry when we  consider only the Riemannian geometry defined by the scalar fields as  the K\"ahler potential, the metric and the Riemann tensor are all symplectic invariant.\\

In  the presence of a  prepotential we have the simpler formula\footnote{$F_{I\cdots K}=\partial_I\cdots\partial_K F$.} 
\begin{align}
\cW_{\alpha \beta\gamma} &=\ii F_{IJK}\frac{\partial Z^I}{\partial z^\alpha}\frac{\partial Z^J}{\partial z^\beta}\frac{\partial Z^K}{\partial z^\gamma},\\
\cN_{IJ} &= \bar F_{IJ}+2 \ii \frac{\Im (F_{IK})\Im(F_{JL}) Z^K Z^L}{\Im(F_{KL})Z^K Z^L}.
\end{align}

 \subsubsection{Minimal  and very special geometry }

{\em Minimal special geometries } correspond to  quadratic prepotential defined with a metric $\eta_{IJ}$ of signature $(1,n)$:
\begin{equation}
F(Z)=-\ii Z^I \eta_{IJ}Z^J,  \quad 
\eta_{IJ}=
 \begin{pmatrix}
1 & \mathrm{0}_{1\times n}\\
\mathrm{0}_{n\times 1} & -\unity_n 
\end{pmatrix},
\end{equation}
called the {\em minimal prepotential}. The corresponding special manifold is $\frac{\SU(1,n)}{\U(1)\SU(n)}$. \\

In a sense, minmal special manifolds are the  simplest kind of special manifolds as the tensor $\cW_{\alpha \beta \gamma}$ identically vanishes leaving the Riemann curvature 
\begin{equation}
R^\alpha{}_{\beta\gamma}{}^\delta=
\delta^\alpha_\beta \delta^\delta_\gamma +\delta^\alpha_\gamma \delta^\delta_\beta. 
\end{equation}

{\em Very special K\"ahler geometries} are characterized by  cubic prepotentials 
\begin{equation}
F(Z)=\ii d_{IJK}\frac{Z^I Z^J Z^K}{Z^0},
\end{equation}
 where $d_{IJK}$ is a real symmetric tensor. \\

Very special K\"ahler geometries  can be obtained by dimensional reduction from $5$ dimensional supergravity theories and  are known to admit  flat potentials. 
 They are also familiar in string theory where they occur in many different compactifications as for example  in toroidal compactifications of the heterotic string with possible Wilson lines \cite{Andrianopoli:1997cm}, in  compactifications of type $II$ string theories on Calabi-Yau threefolds, in compactification of type $II$ string theories  on orientifolds like $K_3\times T^2/\mZ_2$ in the presence of $D3$ and $D7$ branes \cite{Andrianopoli:2003jf,Angelantonj:2003zx}. \\
 
Although   unusual, minimal special geometry  is not incompatible with  string theory. To the best of our knowledge,  there is  so far only one case in which it occurs in string theory \cite{D'Auria:2002tc}. This are the   $\cN=2$ vacua coming from the $\cN=3$  flux compactification  on $T^6/\mZ_2$ studied by  Frey and Polchinski \cite{FP}. \\

In $\cN=2$ compactifications of string theory, it is important for phenomenological reasons to be able to perform a partial  spontaneous supersymmetry breaking to $\cN=1$. As partial supersymmetry breaking is only  possible with symplectic sections that do not admit a prepotential \cite{Ferrara:1995gu},  minimal and very special geometries usually occur in these types of sections that are related to a prepotential only after a symplectic rotation.  \\

A classification of special manifolds was presented  by Cremmer and van Proeyen  in \cite{Cremmer:1984hc}. 

We shall provide some examples to illustrate some subtleties of special geometry.
The first example illustrates that the same manifold can be  endowed with different types of special geometries. This is the content of table \ref{tablesu11}, where we  review the case of the  symmetric space $\frac{\SU(1,1)}{\U(1)}$ which is a K\"ahler manifold of complex dimension one. \\

\begin{table}[htb]
\begin{center}
\begin{tabular}{| l | c | l |}
\hline
Prepotential & Symmetric space \\
\hline
&  \\
$F(X)=-\ii [(X^0)^2-(X^1)^2]$ & $\frac{\SU(1,1)}{\U(1)}$ \\
& \\
\hline
& \\
 $F(X)=\ii \frac{(X^1)^3}{X^0}$ &$\frac{\SU(1,1)}{\U(1)}$  \\
 & \\
\hline
& \\
 $F(X)=-4\sqrt{ X^0(X^1)^3\ }$   & $\frac{\SU(1,1)}{\U(1)}$  \\
& \\
\hline
\end{tabular}
\caption{ In this   table, all three special manifolds correspond to the same coset space $\frac{\SU(1,1)}{\U(1)}$. The first one admits a quadratic prepotential and therefore corresponds to the minimal special geometry. The second one is a very  special manifold as it admits a cubic prepotential. 
 The last one is related to the second one by a  symplectic rotation \cite{deWit:1995jd,Cremmer:1984hc}.\label{tablesu11}} 
 \end{center}
 \end{table}

The second example \cite{deWit:1995jd} present a symplectic section that is not derived from a  prepotential. It is obtained by a symplectic rotation from a minimal special geometry. 
It will also illustrate how a symplectic section  can modify the kinetic terms of the vector fields.

Let us consider the minimal prepotential $F=-\ii Z^0 Z^1$  \footnote{This is related to the usual prepotential of minimal special geometry $F=\ii (Z^0Z^0-Z^1Z^1)=-\ii(Z^0+Z^1)(Z^0-Z^1)$ by the redefinitions 
$Z^0\Longrightarrow Z^0+Z^1$ and $Z^1\Longrightarrow Z^0-Z^1$.
}. The corresponding symplectic section is 
\begin{equation}
v= \begin{pmatrix} Z^0 \\ Z^1\\  - \ii Z^0\\ -\ii Z^1  \end{pmatrix}=
 \begin{pmatrix} 1 \\ z\\  -\ii z \\ -\ii   \end{pmatrix},\quad z=\frac{Z^1}{Z^0}.
\end{equation}
The K\"ahler potential is $\cK=-\log 2(z+\bar z)$.
The kinetic matrix for the vector is  
\begin{equation}
\cN=\begin{pmatrix} -\ii z & 0 \\ 0 & - \frac{\ii}{z} \end{pmatrix} .
\end{equation}

The bosonic part of the $\cN=2$ supergravity coupled to a vector multiplet with a special geometry defined by the previous symplectic section is 

 \begin{equation}
\rme^{-1}\cL_{\text{bosonic}}=\ft12 \mathrm{R}+\frac{\partial_\mu z \partial^\mu \bar z}{(z+\bar z)^2}-\frac{1}{2}\Re \left[  { z (F^{+0}_{\mu\nu})^2+z^{-1} ( F^{+1}_{\mu\nu})^2}\right].
 \end{equation}
 
After a symplectic rotation with 
$
{\cal S}= 
\begin{pmatrix}
1 & 0 & 0 & 0 \\
0 & 0 & 0 &-1 \\
0 & 0 & 1 & 0\\
0 & 1 & 0 & 0 
\end{pmatrix}
$
 we have 
 \begin{equation}
\tilde v={\cal S} v=\begin{pmatrix} 1 \\ \ii \\  -\ii z  \\  z  \end{pmatrix}, \quad
\tilde\cN=\begin{pmatrix}
\ii z & 0 \\ 0 & -\ii z
\end{pmatrix}.
\end{equation}
The new symplectic section cannot be derived from a prepotential as the upper part of the section does not even have a dependence on the scalar field $z$. The bosonic part of the action after the symplectic rotation is

 \begin{equation}
\rme^{-1}\tilde\cL_{\text{bosonic}}=\ft12 \mathrm{R}+\frac{\partial_\mu z \partial^\mu \bar z}{(z+\bar z)^2}-\frac{1}{2}\Re \left[  { z (F^{+0}_{\mu\nu})^2+z ( F^{+1}_{\mu\nu})^2}\right].
 \end{equation}
 
\subsection{Hypermultiplets and quaternionic-K\"ahler geometry}

In four dimensional spacetime with the usual Minkowski signature, an hypermultiplet is composed of four  real scalar fields and two Majorana spinors. As a Majorana spinor can be decomposed into two chiral spinors  of opposite chirality,  one can describe $n_H$ hypermultiplets in terms of $4n_H$ real scalar fields  $q^X$  ($X=1,\cdots, 4 n_H$) and 
$2n_H$ chiral spinors $\zeta^A$ ($A=1,\dots, 2n_H$) of positive chirality  and $2n_H$ chiral spinors $\zeta_A$ of negative chirality.  The  spinors $(\zeta^A, \zeta_A)$ of hypermultiplets are called the {\em hyperini} and the   $4n_H$ real scalar fields $q^X$  are the {\em hyperscalars}. 
The latter can be seen locally as coordinates of a scalar manifold  $\mathrm{M}_H$ which is constrained by $\cN=2$ supersymmetry to be a quaternionic manifold \cite{Bagger:1983tt} \footnote{ For a review of quaterionic geometry see \cite{Galicki:1987ja, Andrianopoli:1997cm,Bergshoeff:2002qk}. In particular, we shall use the conventions of  appendix B of \cite{Bergshoeff:2002qk}. }.
It follows that locally, the $4n_H$ hyperscalar fields can be seen as  $n_H$ quaternions\footnote{The set of quaternions is defined by $\mH=\{ q_0 1 + q_1 \mathrm{i}+q_2 \mathrm{j}+q_3 \mathrm{k} | q_i \in \mR\}$, with the elements of the basis satisfying $\mathrm{i} . \mathrm{j} = \mathrm{k}$, together with all cyclic permutations and $\mathrm{i}^2=\mathrm{j}^2=\mathrm{k}^2=-1$. A quaternion can be represented by  a $2\times2$ matrix as the  imaginary quaternions can be represented by $(i,j,k)=(-\ii\sigma^1,-\ii \sigma^2,-\ii\sigma^3)$ and therefore we can write  $q=q^0 \unity -\ii q^x \sigma^x$. }.
 \\

The supersymmetry transformations (with only leading terms in the fermions) are of the form  
\begin{align}
\delta q^X &= -\ii f^X_{i A} \bar \epsilon^i \zeta^A+\ii f^{XiA}\bar\epsilon_i\zeta_A,\\
\delta \zeta^A &= \frac{\ii}{2} f_X^{Ai}\gamma^\mu \partial_\mu q^X \epsilon_i -\zeta^B \omega_{X}{}_B{}^A [\delta(\epsilon) q^X],\\
\delta \zeta_A &= -\frac{\ii}{2} f_{XAi}\gamma^\mu \partial_\mu q^X \epsilon^i +\omega_{X}{}_A{}^B \zeta_B[\delta(\epsilon) q^X],
\end{align}
where the tensors  $f^X_{i A}, f^{Xi A},f_X^{i A},f_{Xi A}$  are   functions of the scalar fields $q^X$. They are the starting point in determining the geometry of hypermultiplets. Here we have denoted by:
\begin{equation}
\bar \epsilon^i \equiv (\epsilon^i)^T \cC=(\epsilon^i)^\dag \gamma_0 
\end{equation}
the Majorana conjugate of $\epsilon^i$, where $\cC$ is the (unitary and antisymmetric) charge conjugation matrix. For Majorana spinors the 'Majorana conjugate' equals the 'Dirac conjugate', defined by the l.h.s. of the previous equation \cite{LectParis, VanProeyen:1999ni}.\\

As the chiral projections of a given Majorana spinor  are related by charge conjugation and the scalar fields $q^X$ are real we have the following reality  relations 
\begin{equation}
(f_X^{iA})^*=f_{XiA}, \quad (f^{XiA})^*=f^X_{iA},\quad (\omega_X{}_B{}^A)^*=-\omega_X{}_A{}^B.
\end{equation}

Asking the commutator of two supersymmetries  to be a translation imposes that \cite{Rosseel:2004fa}
\begin{equation}
f^X_{iA} f_Y^{jA}+ f^{X jA} f_{YiA}=\delta^X_Y \delta^i_j,  \quad  f^{iA}_X f^X_{jB} =\delta_j^i \delta_B^A.
\end{equation}

This implies in particular that   $f^X_{iA}$  and $ f_X^{iA}$ are  inverses of each other  as $4n_H\times 4n_H$  matrices: 
\begin{align}
f^{iA}_Y f^X_{iA} = \delta_Y {}^X , \quad
 f^{iA}_X f^X_{jB} =\delta_j^i \delta_B^A.
 \end{align}
 
Defining the matrix \cite{Rosseel:2004fa}
 \begin{equation}
 \mC^{AB}=\ft{1}{2}\varepsilon_{ij}f^{XiA}f_X^{jB},
 \end{equation}
 it satisfies  $\mC \mC^*=-\unity_{2n_H}$ and can be used to express the reality condition relating   $f^X_{iA}$ and $f_{XiA}$ as follows
 \begin{equation}
 f^{XiA}=(f^X_{iA})^*=\varepsilon^{ij}\mC^{AB}f^{X}_{jB}.
 \end{equation}
 
 We shall denote $\mC_{AB}$ the  inverse of $\mC^{AB}$, we have 
 \begin{equation}
 f^X_{iA}=(f^{XiA})^*=f^{XjB }\mC_{BA}\epsilon_{ji}.
 \end{equation}

We shall work  with  conventions where 
\begin{equation}
\varepsilon^{ij}=\ii \sigma^2,\quad  \mC^{AB}=\varepsilon^{ij}\otimes \unity_{st}, \quad i,j=1,2, \quad s,t=1,\ldots, n_H.
\end{equation}
In the previous equation, the  indices $A,B=1,\ldots, 2n_H$ has been decomposed into $A\equiv(i,t), B\equiv(j,s)$ where $i,j=1,2$ and $t,s=1,\ldots, n_H$. Note that in this decomposition, the indices    $i,j$ are not  $\mathrm{R}$-symmetry $\SU(2)$ indices as the hyperini are singlets under the $\SU(2)$-part of the $\mathrm{R}$-symmetry \cite{Bergshoeff:2002qk}. \\

The reality conditions imply  that  any linear operator $U$ acting on variables with an  index $A$ belong to $\Gl(n_H, \mH)$ as it satisfies $U^*=\mC U \mC^{-1}$. \\

Finally the supersymmetry algebra also imposes the relation \cite{Andrianopoli:1997cm,Bergshoeff:2002qk}:
 \begin{align}
\covder_X f_Y^{iA} &=\partial_X f_Y^{iA}-   \Gamma_{XY}{}^Z f_Z^{iA}+f_Y^{jA} \omega_{Xj}{}^i + f_Y^{iB} \omega_{XB}{}^A=0,\label{cc}
\end{align}
where $\Gamma_{XY}{}^Z$ is symmetric in the two lower components and  can be interpreted as an affine connection on the scalar manifold. $\omega_{Xj}{}^i$ is an $\SU(2)$-connection and $\omega_{XB}{}^A$  a $\Gl(n_H,\mH)$ connection.  The presence of these connections indicates that there is a  $\SU(2)\times \Gl(n_H,\mH)$ bundle defined on the scalar manifold $\mathrm{M_H}$ of hypermultiplets.
$f_X^{iA}$ is interpreted as the a vielbein on the scalar manifold and the condition \eqref{cc} ensures that it is  covariantly constant with respect to the $\SU(2)\times \Gl(n_H,\mH)$ bundle. 
\\

 The existence of a $\SU(2)\times \Gl(n_H,\mH)$  bundle on the scalar manifold restricts the geometry of the latter.  This can be most easily discussed  in terms of the  {\em holonomy group} of the   the manifold. 
 The { \em holonomy group} of a manifold is defined as the group of transformations generated by parallel transporting all possible vectors around all possible closed curves in  the manifold. 
Under some mild  assumptions, the  holonomy group of a manifold is generated by the Riemann tensor (Ambrose-Singer theorem). 
As the vielbein is covariantly constant when the $\SU(2)$ and $\Gl(n_H,\mH)$  connections are taken into account,  the full Riemann curvature is the sum of a    $\SU(2)$ and a $\Gl(n_H,\mH)$ curvature.
 It follows that  
   the holonomy of the scalar manifold $\mathrm{M}_H$ is   contained  in $ \SU(2)\times \Gl(n_H,\mH)$.
That is, $\mathrm{M}_H$ is a {\em quaternionic manifold} \cite{Bergshoeff:2002qk}.\\

   The reality condition for the veilbein $f_X^{iA}$ can be translated into the property that the  $2\times 2$ matrices $f_X^t\equiv (f_X^t)^i{}_j$ ($A\equiv(t,j)$) satisfy the condition 
\begin{equation}
(f^t_X)^*= \sigma^2 f^t_X \sigma^2.
\end{equation}
This implies that   $f^t$ can be seen as  $n_H$ one-forms with quaternion entries written in the  representation where the quaternionics units are $(\unity_2,-\ii \sigma^x)$  where  $x=1,2,3$ \footnote{Any $2 \times 2$ matrix $q$ satisfying the condition $q^*=\sigma^2 q \sigma^2 $ can be written as a quaternion 
 $q=q^0 \unity -\ii q^x \sigma^x$ with $q^0,q^x\in \mR$ and $\sigma^x$ are the Pauli matrices.}. We can then say that $f^t_X$ is a {\em quaternionic vielbein} as  at each point of the scalar manifold $\mathrm{M_H}$, the $4n_H$ real scalar fields $q^X$ can be organized into $n_H$ quaternions $q^t$ :  
\begin{equation}
q^t=f^t_X q^X. 
\end{equation}

If the $\cN=2$ supersymmetric theory admits an  action, the scalar fields $q^X$ define a sigma model with target space the hyperscalar manifold  $\mathrm{M}_{H}$ which is required to be a Riemannian space with metric $g_{XY}$:

\begin{equation}
  e^{-1}{\cal L}_{\rm hyper}= -\ft12 g_{XY}\nabla _\mu q^X\nabla ^\mu q^Y.
 \label{Lhyper}
\end{equation}
The metric  $g_{XY}$ is  computed from the vielbein as 
\begin{equation}
g_{XY}=f_X^{iA} f_{YiA}.
\end{equation}
 The connection $\Gamma_{XY}{}^Z$, that appears in the covariant condition for the veilbein is the   Levi-Civita connection for the metric $g_{XY}$. The structure group $\Gl(n_H, \mH)$ is then reduced to the subgroup keeping the metric $g_{XY}$ invariant, that is, $\SU(n_H, \mH)= \Sp(n_H)$. 
Using the isomorphism $\SU(2)\simeq \Sp(1)$, we conclude that the  holonomy of the scalar manifold is then  contained in $\Sp(1)\otimes \Sp(n_H)$. Thus, the scalar manifold of hypermultiplets is by definition a {\em quaternionic-K\"ahler manifold}. \\

  We can define a triplet $J^x$ $(x=1,2,3)$ of complex structures :
\begin{equation}
(J^x)_X{}^Y= \ii  f_X^{iA} (\sigma^x)_i{}^j f^Y_{ j A}
\end{equation}

which  satisfy  the multiplication table of  quaternionic units 
\begin{equation}
J^x  J^y = -\delta^{xy}\unity_{4n_H}+\varepsilon^{xyz}J^z,
\end{equation}
and are covariantly constant 
\begin{equation}
\nabla J^x \equiv \nabla^{\mathrm{LC}}J^x+2\varepsilon^{xyz} \omega^y J^z=0,
\end{equation}
where $\nabla^{\mathrm{LC}}$ is the Levi-Civita covariant derivative. 

Any real linear  combination  $J=a_x J^x$  defines  a  complex structure ($J^2=-\unity$) if 
\begin{equation}
 |\!{\abs{\vec a}}\! |^2=(a_1)^2+(a_2)^2+(a_3)^2=1.
\end{equation}
 It follows that at each point of the manifold there is a sphere of complex structures.  Two such complex structures are related by an  $\SU(2)$ rotation. A {\em quaternionic structure} is the space of all these complex structures $a_x J^x$. It is  globally well defined although this is not necessary the case of  an  individual complex structure $J=a_x J^x$ as  the  $\SU(2)$ bundle is  non-trivial.

A quaternionic-K\"ahler manifold  is an Einstein space and 
the quaternionic structure  $J^x_{XY}$ is proportional to the $\SU(2)$ curvature $\cR^x_{XY}
\equiv d \omega^x +\epsilon^{xyz} \omega^y \omega^z$:
\begin{equation}
\mathrm{R}_{XY}=\frac{1}{4n_H}g_{XY}\mathrm{R},\quad 
\cR^x_{XY}=\frac{1}{2}\nu J^x_{XY}, \quad \nu=\frac{1}{4  n_H (n_H+2)}\mathrm{R}.
\end{equation}

In $\cN=2$ supergravity, the constant $\nu$ is proportional to the gravity coupling constant $\nu=-\kappa^2$. We work in units in which $\kappa=1$, that is $\nu=-1$.\\

In flat space, the supersymmetry generators are globally defined and therefore the $\SU(2)$ bundle is trivial and the holonomy is contained in $\Sp(n_H)$ and $\mathrm{M}_H$ is an {\em hyperk\"ahler manifold}. An hyperK\"ahler manifold can be seen as a quaternionic-K\"ahler manifold with a trivial $\Sp(1)$ bundle so that the $\Sp(1)$-curvature vanishes:
\begin{equation}
\cR^x \equiv d \omega^x +\epsilon^{xyz} \omega^y \omega^z=0.
\end{equation}

 The $\Sp(1)\simeq \SU(2)$ and $\Sp(n_H)$ bundles can be understood geometrically as follows: $\SU(2)$ corresponds to the $\mathrm{R}$-symmetry group that rotates the  supersymmetry generators while $\Sp(n_H)$ is the space of linear transformations of $n_H$ hypermulltiplets that  preserve  the metric of the scalar fields. It also acts on the hyperscalars  $q^X$ and the hyperini.\\

 The non-triviality of the $\SU(2)\simeq\Sp(1)$ bundle in $\cN=2$ supergravity  is responsible for the non-existence of FI terms but, ironically, it is also the main instrument to construct $\cN=2$ supergravity potentials that are compatible with $\cN=1$ supergravity with $D$-term endowed with constant FI term as we shall review shortly.

\subsection{Isometries, gauging and scalar potential}
In $\cN=2$ supergravity coupled to vector multiplets and hypermultiplets, the only way to generate a scalar potential is to promote some of the symmetries of the scalar manifold to be local symmetries. 
This implies a choice of the Killing vectors of the scalar manifolds and a choice of vector fields that will be use as gauge fields in the covariant derivatives. \\

In this paper we only consider Abelian gauging of the symmetries of $\mathrm{M}_H$. 
The gauged symmetry is defined by  the  transformation with parameters $\alpha^I$:
\begin{equation}
  \delta _G q^X= -g\alpha ^I k_I^X,
 \label{gaugetransf}
\end{equation}
where  $k_I^X$ are the Killing vectors that we will gauge with the vector fields $W_\mu^I$. 
To gauge a symmetry all the derivatives of the hyperscalars have to be extended to covariant derivatives. 
The gauge field is taken from the vector multiplets:
\begin{equation}
\nabla_\mu q^X = \partial_\mu q^X +  g W_\mu^I k_I^X.
\label{covder}
\end{equation}

A gauging generates  a deformation of the supersymmetry transformations by fermionic shifts $S^{ij}$, $\cN^\alpha_{ij}$ and $\cN^{iA}$:
\begin{align}
\delta \psi^i_\mu & = \cdots -g\gamma_\mu S^{ij}\epsilon_j,\\
\delta \lambda_i^\alpha &=  \cdots +g N^\alpha_{ij}\epsilon^j,\\ 
\delta \zeta^A &= \cdots +g \cN^{i A}\varepsilon_{ij} \epsilon^j.
\end{align} 

These fermionic shifts also contribute to   fermionic mass terms in the Lagrangian: 
\begin{equation}
e^{-1}\cL_{\psi \text{mass}}=-gS^{ij}\bar\psi_{\mu i}\gamma^{\mu\nu}\psi_{\nu j}
+\frac{1}{2} g g_{\alpha\bar\beta} N_{ij}^{\alpha} \bar\lambda^{\bar\beta i}\gamma^\mu\psi^j_\mu
+2g\cN^{iA}\varepsilon_{ij}\bar\zeta_A\gamma^\mu\psi^j_\mu+\text{h.c}
\end{equation}

Finally, in order to preserve supersymmetry of the action, the previous mass terms should be  balanced by the  following scalar potential quadratic in the fermionic shifts\footnote{This is because under the  supersymmetry transformations, $\cL_{\psi\text{mass}}-e  \cV$ is the only sector of the Lagrangian that generates non-derivative fermionic terms of order $g^2$ and linear in the gravitini $\psi_{\mu i}$. We recall that  the determinant of the vielbein transforms as  $\delta e=\ft{1}{2} e \bar\epsilon^i \gamma^\mu \psi_{\mu i}+\text{h.c}$. }:
\begin{eqnarray}
  g^{-2}{\cal V}&=&\ft12g_{\alpha \bar \beta }N^{\alpha
}_{ij}N^{\bar \beta
  ij}+2{\cal N}^{iA}{\cal N}_{iA}
  -6S^{ij}S_{ij}.
\end{eqnarray}

The  invariance of the action under supersymmetric transformations implies that the fermionic shifts are given by :
\begin{eqnarray}
S^{ij} \equiv -{\cal P}_I^{ij} X^I, \quad
N^\alpha _{ij} \equiv   \varepsilon _{ij}k^ \alpha _I \bar X^I -2{\cal
P}_{I ij}\bar {\cal D}_{\bar \beta }\bar X^I
  g^{\alpha \bar \beta } ,
 \qquad {\cal N}^{iA}\equiv -\rmi f^{iA}_Xk_I^X\bar X^I,\label{shifts}
\end{eqnarray}
where $\cP^{ij}_I=\cP^x_{I}(\ii\sigma^x)^{ij}$ and $\cP_{Iij}=\cP^x_{I}(\ii\sigma^x)_{ij}$ are the {\em moment maps} \cite{Galicki:1987ja,Andrianopoli:1997cm,Bergshoeff:2002qk} related to the Killing vectors $k^X_I$ of the quaternionic-K\"ahler manifold and $k^\alpha_I$ are the Killing vectors of the special manifold. 
 Taking into account the relations \eqref{shifts}, the scalar potential reads 
\begin{eqnarray}
  g^{-2}{\cal V}&=&\ft12g_{\alpha \bar \beta }N^{\alpha
}_{ij}N^{\bar \beta
  ij}+2{\cal N}^{iA}{\cal N}_{iA}
  -6S^{ij}S_{ij}, \nonumber\\
&=&(4g_{\alpha \bar \beta}k^\alpha_I k^{\bar \beta}_J +2 g_{XY} k^X_I
k^Y_J) \bar X^I X^J+\left ( U^{IJ}-3 \bar X^I X^J \right ) {\cal P}^x_I {\cal
P}^x_J.\label{N2scalarpotential}
\end{eqnarray}
We also have 
\begin{equation}
U^{I  J } \equiv g^{\alpha \bar \beta}f^I _\alpha f^J _{\bar\beta} = -\ft{1}{2}(\Im \, \cN)^{-1|I  J }-\bar X^I  X^J ,\quad f_\alpha^I  \equiv (\partial_\alpha  +\ft{1}{2}\partial_\alpha \cK)X^I,\quad X^I=\rme^{\frac{\cK}{2}}Z^I.
\end{equation}
As  the scalar fields of vector multiplets transform in the adjont representation of the gauge group,  the Killing vectors $k^\alpha_J, k^{\bar\alpha_J}$ of the special manifold vanish  for Abelian gauging. In particular, the sector  $(4g_{\alpha \bar \beta}k^\alpha_I k^{\bar \beta}_J  \bar X^I X^J)$ of the scalar potential is not present for Abelian gauging.

\subsubsection{Moment map and Fayet-Iliopoulos terms in $\cN=2$ supergravity}
The  moment map $\cP^x_I$  appearing in the fermionic shifts and the scalar potentials is defined as a solution of the following equation \cite{Galicki:1987ja,Andrianopoli:1997cm,Bergshoeff:2002qk}:
\begin{equation}
\frac{1}{2} k_I ^X J^x_{XY}\rmd q^Y=\nabla_Y
\cP_I^x, \quad \text{where }
\nabla
\cP_I^x= \rmd\cP^x_I +2\varepsilon^{xyz}\omega^{y}\cP^z_I.
\label{dP}
\end{equation}

In  rigid $\cN=2$ supersymmetry, the $\SU(2)$ curvature of the quaternionic manifold is trivial. The equation defining the moment map is then 

\begin{equation}
\frac{1}{2} k^X_I J^x_{XY}= \partial_Y  \cP^x_I .
\end{equation}
In flat space, as the moment map is covered by a total derivative in its defining equation, it is determined only modulo an arbitrary triplet  $\xi^x$ of real constants :
\begin{equation}
\cP^x_I \sim \cP^x_I+\xi^x_I.
\end{equation}
These constants $\xi^x$ are the  $\cN=2$ Fayet-Iliopoulos terms. \\

In local $\cN=2$ supersymetry, due to the non-trivial $\rSU(2)$ connection, the triplet moment maps
cannot be shifted by arbitrary constants and there  are only Fayet-Iliopoulos terms when $n_H=0$, that is when there are no hypermultiplets. 

Indeed, as  the $\SU(2)$ bundle is non-trivial, the moment maps are uniquely defined by  
\begin{equation}
 4n_H{\cal P}^x_I =-J^x_{YZ} \nabla^Z k^Y_I,
\end{equation}
thanks to the identity  satisfied by any moment map $\cP^x_I$ \cite{D'Auria:flows}:
\begin{equation}
\nabla^u \nabla_u \cP^x_I= 2 n_H \cP^x_I.
\end{equation}

The uniqueness of $\cP^x_I$ implies in particular that the following equation 
\begin{equation}
\nabla A^x=0,
\end{equation}
 has no nontrivial solutions. Otherwise, $\cP^x_I+A^x$ would be another solutions of equation  \eqref{dP}.
Moreover, if there is a non-trivial solution, the  integrability condition  $[\nabla_u,\nabla_v] A^x=0$ implies that the $\SU(2)$ curvature vanishes and therefore that the $\SU(2)$ bundle is trivial. This is clearly not the case for a  quaternionic-K\"ahler manifold. \\

The moment map can also be described in another way. A Killing vector
preserves the connection $\omega^x$ and K{\"a}hler two forms $J^x$ only
modulo an $\rSU(2)$ rotation. Denoting   by $\cL_I $   the  Lie derivative 
with respect to $k_I$, we have
\begin{equation}
\cL_I \omega^x  = -\ft12 \nabla r^x_I, 
\end{equation}
or in terms of $J^x$
\begin{equation}
 \cL_I
J^x=\varepsilon^{xyz}r^y_I J^z, \label{Liederinr}
\end{equation}

Here $r^x_I$ is known as an $\rSU(2)$ {\em compensator}. The
$\rSU(2)$-bundle of a quaternionic manifold is non-trivial and therefore
it is  impossible to get rid of the compensator $r_I^x$ by a
redefinition of the $\rSU(2)$ connections.\footnote{Again, this is in
contrast with $\cN=2$ rigid supersymmetry, since hyper-K{\"a}hler manifolds
have a trivial $\rSU(2)$ bundle, and therefore no compensator.} The
moment map can be expressed in terms of the triplet of connections
$\omega^x$ and the compensator $r^x_I$ in the following way
\cite{Galicki:1987ja}:
\begin{equation}
\cP^x_I =\ft12 r^x_I+ \iota_I \omega^x,
\label{compensator}
\end{equation}
where $\iota_I$ is an interior derivative with respect to $k_I^X$ ($\iota_I \omega^x=k^Y_I \omega^x_Y$).

\subsection{$\cN=2$ Supersymmetry transformations}

The supersymmetry transformations will be used later on to obtain the BPS equations for the string configuration.\\

The supersymmetry transformations  involve the geometrical objects that we just discussed: the moment map, the Killing vectors and the metric of the scalar manifold.

The bosons transform as
\begin{eqnarray}
 \delta e_\mu ^a& = & \ft12\bar \epsilon ^i\gamma ^a\psi _{\mu i}+\ft12\bar \epsilon _i\gamma ^a\psi _\mu ^i, \nonumber\\
\delta W_\mu ^I&=&\ft12({\cal D}_\alpha X^I)\varepsilon^{ij} \bar
\epsilon _i\gamma_\mu\lambda _j^\alpha +\ft12({\cal D}_{\bar \alpha} \bar
X^I)\varepsilon_{ij} \bar \epsilon ^i\gamma_\mu\lambda ^{\bar \alpha j}+
\varepsilon ^{ij}\bar \epsilon _i\psi _{\mu j} X^I+ \varepsilon
_{ij}\bar \epsilon ^i\psi^j _\mu \bar  X^I, \nonumber\\
 \delta z^\alpha  & = & \ft12\bar \epsilon ^i\lambda _i^\alpha , \nonumber\\
\delta q^X & = & -\rmi f^X_{iA}\bar \epsilon ^i\zeta ^A+\rmi f^{XiA}\bar
\epsilon _i\zeta _A.
 \label{susybosons}
\end{eqnarray}

For a bosonic configuration, the $N=2$ supersymmetry transformations of
the left-handed fermionic fields are:
\begin{eqnarray}
 \delta \psi _\mu ^i & =  & \nabla _\mu(\omega ) \epsilon^i
  +\ft14 \gamma^{\rho \sigma }T^-_{\rho \sigma
}\varepsilon^{ij}\gamma_\mu\epsilon_j 
-g\gamma_\mu S^{ij}\epsilon _j
,\nonumber\\
 \delta \lambda ^\alpha_ i&=&{\slashed \nabla }z^\alpha \epsilon _i
-\ft12g^{\alpha \bar \beta }{\cal D}_{\bar \beta }
 \bar X^I\Im{\cal N}_{IJ}F^{-J}_{\mu \nu }\gamma ^{\mu \nu }\varepsilon
_{ij}\epsilon^j
+ gN^\alpha _{ij}\epsilon ^j,\nonumber\\
 \delta \zeta ^A&=&\ft12\rmi f^{Ai}_X\slashed{\nabla }q^X\epsilon_i
+g{\cal N}^{iA}\varepsilon _{ij}\epsilon ^j.\label{susytransformations}
\end{eqnarray}

The fermionic shifts ($S^{ij}, N^\alpha_{ij}$ and ${\cal N}^{iA}$) are given in equation \eqref{shifts}. The  associated scalar potential is presented in equation \eqref{N2scalarpotential}.
The covariant derivatives are
\begin{eqnarray}
 \nabla _\mu(\omega) \epsilon^i &\equiv &
 \left( \partial_\mu + \ft14 \omega_\mu{}^{ab} \gamma _{ab}\right) \epsilon ^i
 +\ft12\rmi A_\mu \epsilon^i+ V_{\mu j}{}^i\epsilon ^j,\nonumber\\
 \nabla _\mu z^\alpha &= & \partial _\mu z^\alpha +gW_\mu ^I k_I^\alpha,
\nonumber\\
 \nabla _\mu q^X & = & \partial _\mu q^X+gW_\mu ^Ik_I^X.
\end{eqnarray}
We included here the effect of a gauging in the vector multiplet sector
by the Killing vector $k_I^\alpha $ describing the transformations under
the gauge symmetry of the vector multiplet scalar similar to the
definition of $k_I^X$ for the hypermultiplet
scalars. \\

The $\SU(2)$ connection $V_{\mu i}{}^j$ is related to the
quaternionic-K{\"a}hler $SU(2)$ connection  and gets  a  contribution from the moment map when   isometries  of the quaternionic-K{\"a}hler manifold have been gauged:
\begin{equation}
 V_{\mu i}{}^j =\partial_\mu q^X \omega_{Xi}{}^j + gW_\mu^I
 {\cal P}_{Ii}{}^j.
\end{equation}
$A_\mu $ are the components of the one-form gauge field of the K{\"a}hler
$\U(1)$:
\begin{equation}
  A=-\frac{\ii}{2} \left( \partial _\alpha {\cal K}\rmd z^\alpha
  -\partial _{\bar \alpha }{\cal K}\rmd \bar z^{\bar \alpha }\right) .
\end{equation}
In the case of gauging in the vector multiplet sector, this is modified
by a scalar moment map similar to the $\SU(2)$ connection. The dressed
graviphoton is given by
\begin{equation}
T^-_{\mu\nu}=F^{-I}_{\mu\nu}\Im{\cal N}_{IJ}X^J.
\label{Graviphoton}
\end{equation}

\subsection{$\cN=1$ supergravity}

The matter content of $\cN=1$ supergravity coupled to $n_G$ gauge multiplets and $n_C$ chiral multiplets is   presented in table \ref{N1Matter}. For a review on $\cN=1$ supergravity see \cite{Binetruy:2004hh} . We work with  Majorana spinors  that we split into their chiral components ($\psi_{\mu L},\psi_{\mu R}, \lambda^\alpha,  \lambda_\alpha, \chi^i, \chi_i$),
For  $\lambda$ and $\chi$, the chirality is given by the position of the index  : 
\begin{table}[h]
\begin{center}
\begin{tabular}{|l|l | l | l|}
\hline
& & & \\
& vielbein & $e_\mu^a$ & \\
Gravity multiplet &  &  &$\mu,a=0,1,2,3$\\
& graviton & $\psi_{L\mu},\psi_{R\mu}$ & \\
& & & \\
\hline
& & & \\
& gauge field & $W_\mu^\alpha$ & \\
Gauge multiplets & & & $\alpha=1,\cdots, n_G$ \\
& gaugino & $\lambda^\alpha,\lambda_\alpha$ & \\
& & & \\
\hline
& & & \\
 & scalars of chiral multiplets &  $\phi^i$ &  \\
Chiral multiplets &  &  & $i=1,\cdots, n_C$ \\
& chiral fermions & $\chi^i,\chi_i$ &   \\
& & & \\
\hline
\end{tabular}
\caption{Matter content of $\cN=1$ supergravity coupled to $n_G$ gauge multiplets and $n_C$ chiral multiplets. 
\label{N1Matter}}
\end{center}
\end{table}  

\begin{align}
\psi_{\mu L} &=\ft12 (1+\gamma_5)\psi_\mu, \quad \psi_{\mu R}=\ft12 (1-\gamma_5)\psi_\mu,\nonumber\\  
\lambda^\alpha& =\ft12 (1+\gamma_5)\lambda^\alpha, \quad  \lambda_\alpha=\ft12 (1-\gamma_5)\lambda_\alpha,\nonumber\\
\chi^i &=\ft12 (1+\gamma_5)\chi^i,\quad \chi_i=\ft12 (1-\gamma_5)\chi_i.
\end{align}

The scalar manifold of chiral multiplet is a K\"ahler-Hodge manifold \cite{Bagger}.
The action is completely determined by a K\"ahler potential $\cK(\phi,\phi^*)$, an holomorphic matrix  $f_{\alpha\beta}(\phi)$ for the kinetic terms of the gauge field, a superpotential $\cW(\phi)$, the gauging and the FI terms:

\begin{align}
\rme^{-1}\cL_{\text{bosonic}} & = 
\frac{1}{2}\mathrm{R}
-\frac{1}{4}\Re (f_{\alpha\beta})F^\alpha_{\mu\nu}F^{\beta\mu\nu}
+\rme^{-1}\frac{1}{8} \Im f_{\alpha\beta}\varepsilon_{\mu\nu\rho\sigma}F^{\alpha\mu\nu}F^{\beta\rho\sigma}
+g_{i\bar j}(\nabla_\mu \phi^i)(\nabla^\mu \bar\phi^{\bar j})
-V,
\end{align}
where $F^\alpha_{\mu\nu}=2\partial_{[\mu} W^\alpha_{\nu]}$, $g_{i\bar j}=\partial_i\partial_{\bar j}\cK(\phi,\bar \phi)$ and 
$\nabla_\mu  \phi^i = \partial_\mu \phi^i +g k^i_\alpha W^\alpha_\mu$. The scalar potential $V$ is the sum of a $F$-term potential that depends on the superpotential and a $D$-term potential that depends on the gauging through the moment map $\cP_\alpha$: 
\begin{equation}
V=V_F+V_D= e^{\cK}\left( \cD_i  W g^{i\bar j} \cD_{\bar j} \bar W -3 W \bar W \right)+\frac{1}{2}(\Re f)_{\alpha\beta}D^\alpha D^\beta,
\end{equation}
where $\cD_i W=\partial_i W+ \partial_i \cK W$ and $g^{i\bar j}$ is the inverse of $g_{i\bar j}$.
The $D$-terms $D^\alpha$ are defined by $D^\alpha= (\Re f)^{{-1}\, \alpha\beta}\cP_\beta$, where $\cP_\alpha$ is the moment map related to the Killing vector $k^i_\alpha$

\begin{equation}
\partial_{  i} \cP_\alpha= -\ii g \bar k_{\alpha}^{\bar  j}\, g_{ i\bar j}.
\end{equation}

In this equation, the moment map appears covered by a derivative, therefore it is  only defined modulo a constant shift $\xi_\alpha$ that corresponds to  a constant Fayet-Iliopoulos  (FI) term 
\begin{equation}
\cP_\alpha\sim \cP_\alpha +\xi_\alpha.
\end{equation}

The $\cN=1$ supergravity supersymmetry transformations are 

\begin{align}
\delta e^a_\mu& = \ft12 \ii \bar \epsilon \gamma^a \psi_\mu  , \quad \delta W^\alpha_\mu=-\ft12 \bar \epsilon \gamma_\mu \lambda^\alpha, \quad \delta \phi^i=\bar \epsilon_L \chi^i \nonumber\\
\delta \psi_{\mu L} &=  \left( {\partial_\mu + \ft{1}{4}\omega_{\mu}^{ ab}\gamma_{ab} +\ii \ft{1}{2} A_\mu^B 
}\right)
\epsilon_L+\ft12  m \gamma_\mu \epsilon_R ,\nonumber\\
\delta \lambda^\alpha &= \ft{1}{4}\gamma^{\mu\nu}  F^\alpha_{\mu\nu}\epsilon  +\ii \ft{1}{2}\gamma_5 D^\alpha \epsilon, \nonumber\\
\delta \chi_i & = \ft{1}{2}\slash\!\!\!\!\nabla\phi_i \epsilon_R -  \ft{1}{2} m_i \epsilon_L,\label{N=1Susy}
\end{align}
where 
$A^B_\mu=\ft{1}{2}[(\partial_i \cK) \partial_\mu \phi^i-(\partial_{\bar i}\cK) \partial_\mu \bar \phi_{\bar i}]+W_\mu^\alpha \cP_\alpha $, $m=e^{\ft12 \cK} W$, $m_i=\cD_i m=e^{\ft12 \cK}\cD_i W$.

\newpage
\section{Getting $\cN=1$ FI terms from $\cN=2$ supergravity}

We have seen in the previous section that there are no constant Fayet-Iliopoulos terms in $\cN=2$ supergravity in presence of hypermultiplets. This is due to the nontrivial $\SU(2)$ bundle of the quaternionic-K\"ahler manifold. \\

As we have seen, in $\cN=2$ supergravity coupled to vector and hypermultiplets, the only way to deform  the action is to gauge the theory. 
A gauging in $\cN=2$ supergravity will generate a scalar potential. 
The latter might be interpreted as a $D$-term potential endowed with a constant FI term after  a truncation of the theory. Moreover, the truncation could correspond to a vacuum of the $\cN=2$ supergravity.

\subsection{Consistent reduction of supersymmetry }             %
It is often important to reduce the number of supersymmetries of a given theory. 
This is usually motivated by phenomenology as one would like in general to have the minimal amount of supersymmetry ($\cN=1$). 
They are many ways to realize a  reduction of the number of supersymmetries of a given theory. 
One can break supersymmetry directly by introducing by hand supersymmetry breaking terms in the Lagrangian.  Two other methods that appear naturally in the study of string inspired supergravity theories are   {\em spontaneous supersymmetry breaking} and {\em consistent reduction of supersymmetry} \cite{Andrianopoli:2001zh,Andrianopoli:2001gm,D'Auria:2005yg}. \\

In supergravity theories coming from string theory, spontaneous supersymmetry breakings arrive  naturally from compactification with fluxes and/or torsion. 
In such a case, one ends up with  gauged supergravity theories (or with a superpotential in $\cN=1$ supergravity) in which some of  the gravitini  become massive with masses coming from the fermionic shifts associated with the scalar potential. 
Consistent truncations, on the other hand do not require any fermionic masses,  they consist of a   reduction of the number of fields of the theory (including some massless fields) while respecting the equations of motion of the theory. It is then possible to get rid of some gravitini and supersymmetry generators. Although  they might appear quite artificial at first look from a purely supergravity point of view,  consistent reductions are naturally realized in string theory, for example by  models containing  orbifolds and/or orientifolds.  Moreover some spontaneous supersymmetry breaking are also consistent reduction in the sense that all solutions of the equations of motion of the reduced theory are also solutions of the mother theory.\\

We shall now review the difference between a spontaneous supersymmetry breaking and a consistent truncation in the eyes of the equations of motion of the mother theory.
In the context of $\cN$-extended supergravity, spontaneous supersymmetry breaking consists in  giving mass to $(\cN-\cN')$-gravitini ($\cN'<\cN$)  leaving a  $\cN'$  extended supergravity theory at energy well below the scale fixed by the gravitini masses. The effective theory defined after a spontaneous supersymmetry breaking admits a Lagrangian $\cL'$ depending of a subset $\{\phi^r\}$ of the fields $\{\phi\}$ of the mother theory  $\cL$  such that  
\begin{equation}
\cL(\phi)= \cL'(\phi^{r})+O(M^4), \quad \{\phi^r\}\subset\{\phi\},
\end{equation}

where $M$ is the energy scale of spontaneous supersymmetry breaking (the mass of the gravitini).  The fields of the reduced theory are those that are light with respect to the energy scale defined by the massive gravitini.
The solutions of the equations of motion of $\cL'$ are solutions of the mother theory given by the Lagrangian $\cL$  modulo terms only relevant  at the energy scale fixed by the masses of the gravitini:
\begin{equation}
\text{Spontaneous supersymmetry breaking}\quad\quad\frac{\delta \cL'}{\delta \phi^r}=0\Longrightarrow \frac{\delta \cL}{\delta \phi}=0+O(M^4).
\end{equation}
The same goes for the supersymmetry transformations of $\cL'$, they corresponds  to those of $\cL$ only modulo terms that are supposed to vanish at low energy with respect to the scale of supersymmetry breaking. \\

In a consistent truncation, any solution of the  equations of motion of the reduced theory correspond to a solution of the mother theory.   If $\{\phi\}=\{\phi^r\}\cup \{\phi^T\}$ are the fields  of the mother theory $\cL$,  where $\{\phi^r\}$ are those that survive the  truncation to the reduced theory  $\cL^r$  and $\{\phi^T\}$ are the truncated field, the truncation ansatz is simply $\phi^T=0$. A consistent truncation is such that the truncation ansatz commutes with the equations of motion 
\begin{eqnarray}
\text{Consistent truncation}:&& \frac{\delta }{\delta \phi}\left({\cL |_{\phi^T=0}}\right)  =\left.{\left({\frac{\delta \cL}{\delta \phi}}\right)}\right|_{\phi^T=0},
\end{eqnarray}
and therefore any solutions of the reduced theory is also a solution of the mother theory 
\begin{eqnarray}
\text{Consistent truncation}:\,&\quad&\frac{\delta \cL^r}{\delta \phi^r}=0\Longrightarrow \frac{\delta \cL}{\delta \phi}=0.\text{\; \; \, }
\end{eqnarray}

It is possible that a spontaneous supersymmetry breaking implies a consistent truncation, but this is not necessarily the case. \\

\subsection{$\cN=2\Longrightarrow \cN=1$ consistent truncations }

Consistent truncation of supersymmetry in the context of supergravity is not a trivial task. 
A $\cN$-extended supergravity theory admits $\cN$-gravitini and  cannot be seen in general as a special case of a  $\cN'<\cN$ supergravity theory. Indeed the extra $(\cN-\cN')$ gravitini  generate couplings that are inconsistent with the constraints of a  $\cN'$ extended supergravity.
Let us for example illustrate a few examples of incompatibilities between   $\cN=2$ supergravity theory  and $\cN=1$ supergravity: 
\begin{itemize}

\item 
The graviphoton appearing in the supersymmetry transformation of the gravitini in $\cN=2$ supergravity. 
Such a term is inconsistent with the supersymmetry transformation of $\cN=1$ supergravity. 

\item In $\cN=2$ supergravity, $n_H$ hypermultiplets cannot be seen as $2n_H$ chiral multiplets. Indeed, the scalar fields of $n_H$ hypermultiplets define a quaternionic geometry and in general a quaternionic manifold is not even a complex manifold and therefore does not qualify to describe the geometry of $\cN=1$ chiral multiplets as the latter is supposed to be K\"ahler-Hodge.

\item  $\cN=2$ vector multiplets.  Here the scalar fields of the vector multiplets define a special manifold. A special manifold is a K\"ahler-Hodge manifold and therefore one would expect that $n_V$ $\cN=2$ vector multiplets can be seen as $n_V$ gauge multiplets of $\cN=1$ supergravity together with $n_V$ chiral multiplets. However, this is not in general the case. 
Indeed, the coupling matrix $\cN_{IJ}$ appearing in the kinetic terms of $\cN=2$ gauge fields is not restricted to be an holomorphic function whereas this is mandatory in $\cN=1$ supergravity coupled to gauge fields\footnote{The same argument shows that even in rigid supersymmetry, $\cN=2$ and $\cN=1$ supersymmetry are in general not compatible.}. 
\end{itemize}
The previous points illustrate some of the requirements that a  consistent truncation of $\cN=2$ supergravity to $\cN=1$ supergravity should satisfy:  the graviphoton should vanish, the quaternionic-K\"ahler manifold should reduce to a K\"ahler-Hodge manifold and the coupling matrix $\cN$ should be holomorphic after the truncation. \\

The conditions ensuring a consistent truncation in supergravity have been analyzed  carefully in \cite{Andrianopoli:2001zh,Andrianopoli:2001gm,D'Auria:2005yg}.
 In particular we see that in a $\cN=2\Longrightarrow \cN=1$  consistent  truncation, at least half of the degrees of freedom have to be truncated. In the gravity multiplet, the graviphoton and one gravitino have to be truncated. 
The quaternionic-K\"ahler manifold has to admit a K\"ahler-Hodge submanifold in order to satisfy the constraints of $\cN=1$ supergravity. 
 A hypermultiplet  is fully truncated or reduces to a unique chiral multiplet. A vector multiplet can be completely truncated or reduces to a gauge or a chiral multiplet.

\subsubsection{A dictionary of $\cN=2\Longrightarrow\cN=1$ consistent truncations }

We shall only consider  $\cN=2\Longrightarrow\cN=1$ consistent truncation in which we keep the first gravitino but truncate the second one, that is 
\begin{equation}
\psi_{\mu 2}=\psi_\mu^2=\epsilon_2=\epsilon^2=0.
\end{equation}
After a consistent truncation, the resulting $\cN=1$ supergravity is such that
\begin{subequations}
\begin{align}
\epsilon_L&=\epsilon^1,\quad  &\psi_{\mu L}&= \psi^1_\mu,\\
D^\alpha &=-2g \Im \cN^{-1|\alpha \beta}(\cP^0_\beta+\cP^3_\beta),\quad  
 & \rme^{\ft{1}{2}K} W &=-2gX^\beta(P^1_\beta-\ii\cP^2_\beta),\\
 \lambda^\alpha_L &=-\lambda^\beta_2 \cD_\beta X^\alpha, \quad 
 & f_{\alpha\beta}&= \ii \cN_{\alpha \beta}.
\end{align}\label{truncation}
\end{subequations}
To write these equations we have decomposed the $\cN=2$ vector indices $I$ as   $I \to ( \alpha, \tilde \alpha)$, with $\alpha=1,\cdots,n_G$ running over the $n_G$ 
 gauge multiplets that survive the truncation to the $\cN=1$ theory, and $\tilde \alpha = 0, n_G+1,\cdots,n_V$ over the complementary indices, which label the chiral multiplets of the truncated theory
 coming from $\cN=2$ vector multiplets.\\
 
Among the conditions for a consistent truncation let us mention the following 
\begin{equation}
T_{\mu\nu}=0, \quad \omega^1_X=\omega^2_X=0,
\end{equation}
where $T_{\mu\nu}$ is the graviphoton and $\omega^x_X$ is the $\SU(2)$-connection of the quaternionic-K\"ahler manifold. 

The  scalar manifold of the reduced $\cN=1$ theory is a direct product 
$\mathrm{M^{KH}_V}\otimes \mathrm{M^{KH}_H}$ where $M^{KH}_V$ is the reduced manifold coming from the scalar manifold $\mathrm{M}_V$ of vector multiplets and $\mathrm{M}^{KH}_H$ is a K\"ahler-Hodge submanifold of the quaternionic-K\"ahler manifold $\mathrm{M}_H$:
\begin{equation}
\mathrm{M}=  \mathrm{M}_V\otimes \mathrm{M}_H\Longrightarrow 
 \mathrm{M}_{KH}= \mathrm{M}^{KH}_V \otimes \mathrm{M}^{KH}_H.
\end{equation} 
The $\U(1)$ connection of  $\mathrm{M}^{KH}_H$ is determined by  $\omega^3_X$.

\subsection{$\cN=1$ FI terms from a $\cN=2$  consistent truncation }

As discussed in the introduction, in the context of $\cN=1$ supergravity, constant Fayet-Iliopoulos terms play an important role in cosmology and in the study of topological defects and in particular for constructing potentials admitting cosmic string solutions of Nielsen-Olesen type. \\

However as $\cN=2$ supergravity theories do not admit Fayet-Iliopoulos terms in the presence of hypermultiplets, it is not clear how these constant FI terms of $\cN=1$ supergravity could be obtained from $\cN=2$ supergravity. Indeed, as  we have discussed in the previous section, $\cN=1$ and $\cN=2$ supergravity theories are not necessary compatible. 
However, starting from an $\cN=2$ supergravity theory one can go to $\cN=1$ by spontaneous supersymmetry breaking or by a consistent truncation. \\

If we consider a consistent truncation in which we truncate the second gravitino, we see from the dictionary of the previous section that a constant FI term can be generated by a moment map with a constant $\cP^3$. This would require  defining a gauging of a symmetry of the quaternionic-K\"ahler manifold. \\

Consider a Killing vector $k_1$ of a quaternionic-K\"ahler manifold $M_H$ such that 

\begin{equation}
\cL_{k_1} J^x= \varepsilon^{xyz} r^y_{k_1} J^z,
\end{equation}
with $r^x_{k_1}=\delta^x_3$.
The moment is then 
\begin{equation}
\cP_{k_1}^x=\iota_{k_1}\omega^x +\ft{1}{2}\delta^x_3,
\end{equation}
if we gauge $\xi k_1$ and can have $\iota_{k_1}\omega^x=0$ we have a candidate $\cN=2$ gauging that can be reduced to an $\cN=1$ supergravity with a vanishing superpotential, and a $D$-term endowed with  a constant FI terms because  $D$-term is proportional to $\cP^3_{k_1}$ since there is no gauging along the special manifold:
\begin{equation}
\cD\sim  \xi
\end{equation}

Such a  constant FI term that we obtain in the reduced $\cN=1$ theory is coming from the $\SU(2)$  compensator $r^x_{k_1}$ which is due to the non-triviality of the $\SU(2)$-bundle. Ironically it is  the very same property that forbids the introduction of $\cN=2$ FI terms in supergravity coupled to hypermultiplets!\\

The   mechanism that we have just discussed was first presented  in  \cite{N2DtermString} in a model  where  $\cN=2$ supergravity is coupled to one vector multiplet and one chiral multiplet. After truncation it  yields  a $\cN=1$ supergravity theory coupled to a gauge and a  chiral multiplet and admitting a $D$-term potential endowed with a constant FI term. The FI term was used to construct the first example of a half-BPS cosmic string solutions of $\cN=2$ supergravity.  \\

Using the property of consistent truncations, the construction of \cite{N2DtermString} can be described as the embedding of a $\cN=1$ half-BPS $D$-term cosmic string solution in a $\cN=2$ supergravity theory.  Indeed, any solution of the reduced $\cN=1$ theory is also a solution of the mother $\cN=2$ theory.  One can alternatively  consider the use of consistent truncation in \cite{N2DtermString} as a trick to simplify the $\cN=2$ BPS equations so that they look similar to those of a $\cN=1$ supergravity model.

\subsubsection{A simple example}

We shall now present a realization of the mechanism discussed above in a simple example. In this example, we shall focus on the hypermultiplets. The rest of the paper will emphasize   the role of the special geometry.  For the quaterionic-K\"ahler manifold, we take a quaternionic space of quaternionic dimension one :
\begin{equation}
\frac{\SO(4,1)}{\SO(4)}=\frac{\Sp(1,1)}{\Sp(1)\cdot \Sp(1)},
\end{equation}
with real coordinates $(b^0=\rme^{h}, b^1, b^2, b^3)$ and admitting the metric 
\begin{equation}
\rmd s^2= \frac{1}{(b^0)^2}\left[{(\rmd b^0)^2+\abs{\vec\rmd b}^2}\right].
\end{equation}

The vielbein and $\SU(2)$ connection are given by 
\begin{equation}
f=\frac{1}{\sqrt{2}\, \,b^0}(\rmd b^0\unity_2+\ii{\vec{\rmd b}}\cdot \vec\sigma),
\quad
\vec\omega=-\frac{\vec{\rmd b}}{b^0}.
\end{equation}

The isometry
\begin{equation}\label{k1}
k_1=2 b^1 \frac{\partial }{\partial b^2}- 2b^2 \frac{\partial }{\partial b^1}, 
\end{equation}
rotates the connection $\omega^x$ as 
\begin{equation}
\cL_{k_1} \omega^x=2 \varepsilon^{xyz}\omega^z\delta^y_3=-\nabla \delta^x_3,
\end{equation}
from which we identify the $\SU(2)$-compensator  $r^y_{k_1}=2\delta^x_3$. We can then compute the moment map 
\begin{equation}
\cP_{k_1}=\iota_{k_1}\omega+\ft{1}{2}r_{k_1}=
\ft{1}{b^0}\begin{pmatrix} b^2\\-b^1\\ 0 \end{pmatrix}+
\begin{pmatrix}0 \\ 0\\ 1\end{pmatrix}.
\end{equation}

Along the quaternionic-K\"ahler manifold, the extremum of the scalar potential is  at 
\begin{equation}
b^1=b^2=0,
\end{equation}
leaving the constant moment map 
\begin{equation}
\cP_{k_1}|_{b^1=b^2=0}=\begin{pmatrix}0 \\ 0\\ 1\end{pmatrix}.
\end{equation}

The quaternionic-K\"ahler manifold $\frac{\SO(4,1)}{\SO(4)}$  reduces to the  submanifold  $\frac{\SO(2,1)}{\SO(2)}$
\begin{equation}
\frac{\SO(4,1)}{\SO(4)}\Longrightarrow\frac{\SO(2,1)}{\SO(2)},
\end{equation}
with metric 
\begin{equation}
\rmd s^2=\frac{1}{(b^0)^2}[(b^0)^2+(b^3)^2].
\end{equation}
Defining $\phi= b^3 -\ii b^0$ we have a K\"ahler-Hodge metric 
\begin{equation}
\rmd s^2=\frac{\rmd \phi\bar{\rmd \phi}}{(\Im \phi)^2},
\end{equation}
admitting  the K\"ahler potential $\cK=-2\log [\ii ( \phi-\bar \phi)]$.

We see that the quaternionic-K\"ahler manifold $\frac{\SO(4,1)}{\SO(4)}$ (which is not even a complex  manifold) can be consistently reduced to a K\"ahler-Hodge manifold $\frac{\SO(2,1)}{\SO(2)}$.  \\

In the present case, if we gauge 

\begin{equation}
k=\xi k_1, \label{xi}
\end{equation}
where $k_1$ is given in equation \eqref{k1} we obtain after truncation $(b^1=b^2=0)$  an $\cN=1$ supergravity with vanishing superpotential and a $D$-term endowed with a constant FI term,  using \eqref{truncation} we have:
\begin{equation}
 D= -2g \cP^3=-2g\xi.
\end{equation}

As we see the coefficient $\xi$  in \eqref{xi} determined the value of the constant FI term. This is because   $k_1$ has been normalized  such that the third component of the moment map is equal to one ($\cP^3_{k_1}=1$). 
Since the Killing vector $k_1$ vanishes identically, there is no gauge symmetry in the reduced $\cN=1$ theory. But if we gauge a linear combination $k=\xi k_1 +k_2$ such that  $k_2$ acting on $b^0$ and $b^3$ we can end up with an $\cN=1$ theory with an Abelian gauging and a constant Fayet-Iliopoulos term as we shall discuss in the next section.

\subsubsection{Engineering  cosmic strings in $\cN=2$ supergravity}

Once we  can engineer  a  gauging in $\cN=2$ supergravity yielding FI terms in a reduced  $\cN=1$ supergravity, it is not difficult  to identify  a gauging that generates a $\cN=2$ potential  admitting  cosmic string solutions of the Nielsen-Olesen type. 
What we need is to have a $\U(1)$ compact symmetry surviving the reduction and gauging the Higgs field of the string. \\

Since we work with a consistent truncation, if the reduced theory admits cosmic string solutions, they whould also be solutions of the mother $\cN=2$ supergravity equations of motion. 
In this spirit, we first identify a Killing vector in the reduced $\cN=1$ theory which  generates a compact $\U(1)$ symmetry. We then lift up that  symmetry to the  quaternionic-K\"ahler manifold\footnote{
This is always possible as the sector of the scalar manifold of the $\cN=1$ coming from the quaternionic-K\"ahler manifold is a completely geodesic submanifold of the former. Therefore any symmetry of the reduced manifold can be lifted up to a symmetry of the mother manifold. }.\\

Let us illustrate the point using the previous example. The reduced K\"ahler-Hodge manifold  $\frac{\SO(2,1)}{\SO(2)}$ admits a unique  compact  symmetry 
\begin{equation}
\delta \phi=-g(\phi^2+1).
\end{equation} 

The cosmic string solutions are based on the $D$-term potential 
\begin{equation}
D\propto -2g\left({\frac{\abs{\phi}^2+1}{\Im \phi}-\xi}\right).
\end{equation}
The vacuum is a circle if we take  ($\xi>2$)    \footnote{$\xi<2$ is excluded as $b^0=\rme^h>0$.}
\begin{equation}
D=0\Longrightarrow (b^3)^2+(b^0-\frac{\xi}{2})^2=\frac{\xi^2-4}{2}.
 \end{equation}

We then gauge in the mother $\cN=2$ supergravity theory the Killing vector 
\begin{equation}
k=\xi k_1+ k_2,
\end{equation}
where 
\begin{equation}
k_2= (\phi^2+1)\frac{\partial}{\partial \phi}+\cdots
\end{equation}is an uplift of the compact symmetry of  $\frac{\SO(2,1)}{\SO(2)}$   to the quaternionic-K\"ahler manifold $\frac{\SO(4,1)}{\SO(4)}$:  the dots $(\cdots)$ stand for  terms that vanish in the reduction.\\

In a sense, we can say that the method that we have presented here is a consistent embedding of an $\cN=1$ theory with a $D$-term endowed with a constant FI term into a gauged  $\cN=2$  supergravity. 
The cosmic string solutions of the reduced theory are also valid solutions of the mother theory. 
Thus we also have a consistent embedding of a cosmic string solution of a $\cN=1$ supergravity theory into $\cN=2$ supergravity.

We note that the gauge symmetry of the reduced symmetry admits a unique fixed point  
\begin{equation}
\phi=\ii,
\end{equation}
where the $D$-term is such that 
\begin{equation}
D\propto  2g(\xi-2). 
\end{equation}

We shall see later on that the deficit angle of the cosmic string is proportional to $(\xi-2)$.

\section{The  model }

In this section we describe the model.  We shall gauge a unique Abelian Killing vector of the quaternionic manifold. 
As for any Abelian gauging in $\cN=2$   supergravity, the scalars of the vector multiplet are neutral as they transform in the adjoint representation of the gauged group. \\

We will consider a simple model involving the minimum number of scalar fields required to have a cosmic string with a $\cN=2$ potential bounded from below using the $ST[2,n]$ manifold. 
As we shall see this will require $n\geq 1$ as the two first lines of the symplectic section are related to the graviphoton and for gauging of the graviphoton the scalar potential is not guaranteed to be bounded from below. Moreover, we would like to define a string configuration in $\cN=2$ supergravity which is compatible with $\cN=1$ supergravity. In such a case, the graviphoton has to decouple on the string configuration. \\

On the quaternionic side, we take the scalar manifold to be $\frac{\SO(4,1)}{\SO(4)}$. The same solution can be constructed using any normal quaternionic manifold. However for keeping the geometry as simple as possible we shall restrict ourselves to $\frac{\SO(4,1)}{\SO(4)}$. This will also enable us to  compare our results with those of \cite{N2DtermString}.\\

The choice of the Killing vector is a crucial part of the construction. For the specific quaternionic geometry that we consider it is explained in \cite{N2DtermString}. The treatment for a generic homogeneous   quaternionic  manifold  will be presented in \cite{MboyoToAppear}.

\subsection{ The very special K\"ahler manifold  $\frac{\SU(1,1)}{U(1)}\times \frac{\SO(2,n)}{\SO(2)\times \SO(n)}$ in the Calabi-Visentini section}
We consider the K\"ahler-Hodge  manifold 
\begin{equation}
ST[2,2+n]=\frac{\SU(1,1)}{\U(1)} \times \frac{\SO(2,2+n)}{\SO(2)\times \SO(2+n)},
\end{equation}

We will work in the  Calabi-Visentini basis  defined by the  holomorphic section : 
\begin{align}
v = 
 \begin{pmatrix}
Z^I \\
F_I 
\end{pmatrix},\quad \text{ with  }\quad 
Z^I = 
\begin{pmatrix}
\frac{1}{2}(1+y^2)\\
\ii\frac{1}{2}(1-y^2)\\
y^a 
\end{pmatrix} , 
\quad \text{ and }    F_I = S \eta_{IJ} Z^I,\label{CV}
\end{align}
 where $ a=1,\cdots, n $ and $y^2= y^a y^a$ and $\eta_{IJ}=\left({\begin{smallmatrix} \unity_2 & \\ & -\unity_n\end{smallmatrix}}\right)$. The fields $S$ and $y^a$ parametrize respectively the manifold $\frac{\SU(1,1)}{\U(1)}$ and $\frac{\SO(2,n)}{\SO(2)\times \SO(n)}$.
The Calabi-Visentini basis does not admit a prepotential, but can be rotated to a symplectic section  
which can be obtained from the cubic prepotential 
\begin{equation}
F(S,y)=-\frac{1}{2} S y^a y^a. 
\end{equation}

The  k\"ahler potential for the Calabi-Visentini section is
\begin{align}
\cK &= -\log \left[{\ii (S-\bar S)}\right]-\log\left[{\frac{1}{2}\left({
1-2 \bar y^a y^a +\abs{y^a y^a}^2
}\right)}\right],\\
\intertext{and the coupling  matrix of the vector field }
\cN_{I  J } & = (S-\bar S)\frac{Z_I  \bar Z_J+\bar Z_I  Z_J}{\bar Z^T \eta Z}+\bar S \eta_{I  J }.
\end{align}
The metric is given as usual by the second derivative of the K\"ahler potential \begin{equation}
g_{S\bar S}=\frac{1}{(2\Im S )^2}, \quad 
g_{b \bar c}=
2\frac{(\delta^{b\bar c}
-2 y^b \bar y^{\bar c})}{1-2 \bar y^a y^a +\abs{y^a y^a}^2}
+
4 \frac{\left[{\bar y^b -y^b (\bar y^{\bar a} \bar y^{\bar a} )}\right]
\left[ {y^{\bar c} - \bar y^{\bar c} (y^a  y^a )}\right]}
{\left({1-2 \bar y^a y^a -\abs{y^a y^a}^2}\right)^2}
\end{equation}

In the example we shall consider in this paper, we will  restrict ourselves to the case $n=1$, as it  requires the minimum amount of fields : two complex scalar fields $S$ and $y$ and three vector fields $W_0,W_1, W\equiv W_2$. This specific case is immediately generalized to any $n$. 
In this case the metric for the scalar manifold $\mathrm{M_V}$ and the coupling matrix $\cN$ simplify to:
\begin{equation}
g_{S\bar S}=\frac{1}{(2\Im S )^2}, \qquad g_{y \bar y}= \frac{2}{(1-y \bar y)^2} \qquad \cN =  
 \begin{pmatrix}
S \unity_2 & \\ 
& -\bar S
\end{pmatrix}
\end{equation}

\subsection{The quaternionic manifold $\frac{\SO(4,1)}{\SO(4)}$}

The quaternionic manifold of quaternionic dimension one 
\begin{equation}
\frac{\SO(4,1)}{\SO(4)},
\end{equation}
has a very simple quaternionic stucture which can be derived from the vielbein 
\begin{equation}
f= \frac{1}{\sqrt{2}} \left( {dh \unity +\ii \rme^{-h} \rmd b^x \sigma^x }\right),
\end{equation}
where $x=1,2,3$ and $h$, $b^x$ are real fields. 
Its metric and $\SU(2)$-connection are respectively 
\begin{align} 
ds^2 &= (\rmd h)^2+\rme^{-2h}\left[ (\rmd b_1)^2+(\rmd b_2)^2+(\rmd b_3)^2  \right],
\label{hypermetric}\\
\omega^x & = -\frac{1}{2}\rme^{-h} \rmd b^x.
\end{align}

\subsection{ Killing vector and  moment map}
\
We will consider  the same Abelian gauging as in \cite{N2DtermString}. 
With our choice of special geometry, on the submanifold $y=0$, the graviphoton, (\ref{Graviphoton}), depends only on  $W_0$ and $W_1$. 
As we would like to put the graviphoton to zero on the string configuration we shall gauge the vector  $W^2_\mu$. 

The $\U(1)$ symmetry that we gauge  is \cite{N2DtermString} : 
\begin{equation}
k= \begin{pmatrix}
4 b_3 \\
4 b_1 b_3 \\
4 b_2 b_3 \\
2\left[b_3^2-\rme^{2h}+1-b_1^2-b_2^2\right]
\end{pmatrix}+\xi 
\begin{pmatrix}
0\\
-2b_2\\
2b_1 \\
0
\end{pmatrix},
\label{killing}
\end{equation}
where we have arranged the tangent vector  in the order $(\frac{\partial  }{\partial h},\frac{\partial}{\partial b_1},\frac{\partial}{\partial b_2},\frac{\partial}{\partial b_3})$. 
Although the  previous Killing vector seems  complicated at first sight it is defined in a precise and simple way on any symmetric normal quaternionic manifold  using a solvable parametrization of the quaternionic manifold \cite{MboyoToAppear}.  
The  moment map corresponding to $k$ reads
\begin{equation}
\cP^x =
\begin{pmatrix}
-2 b_2 -2 b_1 b_3 \rme^{-h} \\
 2  b_1 -2 b_2 b_3 \rme^{-h}\\
-\rme^{-h}\left[{(b_3)^2+1-(b_1)^2-(b_2)^2}\right]-\rme^{h}
\end{pmatrix} 
+\xi 
\begin{pmatrix}
 \rme^{-h} \,  b_2 \\
- \rme^{-h}\,    b_1 \\
1
\end{pmatrix} .
\label{momentmap}
\end{equation}

\subsection{The scalar potential}

In the Calabi Visentini symplectic section we have for any $n$:
 \begin{equation} 
U^{I  J }-3 \bar X^I  X^J =- \frac{1}{\ii  (S-\bar S) }
\eta^{I  J }. 
\end{equation}
Since $Im S<0$, it follows that the scalar potential is always positive and bounded from below in the Calabi-Visentini basis, provided that we gauge the vector $W_I$  with $\eta_{II}$  negative. This corresponds to the gauge field associated with the coordinate $\mathbf{y}$.\\

We shall gauge a unique Killing vector $k^X$ of the hypermanifold. With our choice the  scalar potential is then 
\begin{equation}
\cV = 4\rme^{-\rho} k^2 \frac{ y \bar y}{(1-  y\bar  y)^2}  + 2\rme^{-\rho} \cP^x\cP^x,
\end{equation}
where $k^2=g_{XY}k^X k^Y$. 

The previous scalar potential has the following  properties :  
\begin{enumerate}
\item {\em The scalar potential $\cV$ is bounded from below }
\begin{equation}
\cV\geq 0,
\end{equation}
this is in sharp contrast to the case of the  minimal special geometry where the scalar potential was not bounded from below and could be  positive, negative or vanish  depending on the value of the scalar fields. 
\item {\em $y=0$ is a critical point of the  scalar potential $\cV$}:
\begin{equation}
\frac{\partial \cV}{\partial y}|_{y=0}  =0.
\end{equation}

\item The scalar potential $\cV$ has a runaway behaviour in  the dilaton field  $\rho$ :
\begin{equation}\label{runAway}
\cV\propto \rme^{-\rho} .
\end{equation}
\end{enumerate}

\section{A half-BPS cosmic string solution}

 In this section we will present a 1/2-BPS cosmic string solution of the full $\cN=2$ supergravity action.
First we shall study the Minkowski vacua of the scalar potential. 
Next we will specify the field configuration characterizing the consistent reduction, and finally we will compute the BPS equations for the string configuration and analyze some of the string properties. 

\subsection{Minkowski vacua of the scalar potential}
The scalar potential is a sum of squares.
It is easy to compute all its  Minkowski vacua by looking at the zeroes of the  different  terms: 
\begin{equation}
\cV=0 \Longrightarrow (k=0 \text{ or  }  y=0) \text{ and  } \mathbf{\cP^x \cP^x=0}.
\end{equation}

To study the scalar potential, it  is useful to  introduce the following definitions: 
\begin{align}
\Phi  &=-b_3 +\ii\, \rme^h \quad , \quad  \tilde\Phi= b_1+\ii\,  b_2.
\intertext{ which yield  }
\cP^x \cP^x &= \frac{4}{(\Im  \, \Phi)^2} \left|{ \Phi -\frac{\xi}{2} \, \ii }\right|^2|\tilde\Phi|^2+\left(\frac{\abs{\tilde\Phi}^2}{\Im \, \Phi} -\frac{\abs{\Phi}^2+1}{\Im \, \Phi} +\xi\right)^2\\
\label{momentmap2}
\end{align}

It is then easy to find    
  \begin{equation}
\mathbf{\cP^x \cP^x}=0 \Longrightarrow 
\begin{cases}
\text{Case I} &: \Phi =\frac{\xi}{2}\, \ii, \qquad  \abs{\tilde \Phi}^2 =1-\frac{\xi^2}{4}  ,\quad (0<\xi<2), \\
\\
\text{Case II}&:\frac{\abs{\Phi}^2+1}{\Im  \, \Phi}=\xi, \qquad   \tilde\Phi =0 , \quad  (2\leq  \xi).
\end{cases}
\end{equation}

 $k=0$ has a unique solution given by the origin of the quaternionic manifold :
\begin{equation}
k =0 \Longrightarrow b_3=b_2=b_1=h=0\iff \Phi=\ii \quad \text{ and } \quad \tilde\Phi=0.
\end{equation}

Putting things together we have a Minkowski vacuum for each value of $0<\xi$:

 \begin{table}[htb]
\begin{center}
\begin{tabular}{|lll||r|}
\hline 
\multicolumn{4}{|c|}{Minkowski vacua }\\
\hline
\hline
& & & \\
\multicolumn{3}{|c||}{No Minkowski vacuum} & $\xi\leq 0$\\
 & & & \\
\hline 
 & & & \\
$y=0$, & $\Phi=\frac{\xi}{2}\ii $, & $\abs{\tilde\Phi}^2=1-\frac{\xi^2}{4}$ & $0<\xi<2$ \\
 & & & \\
\hline
 & & & \\
& $\Phi=\ii $, & $\tilde\Phi=0$ & $2=\xi$ \\
 & & & \\
\hline 
 & & & \\
$y=0$, & $\frac{\abs{\Phi}^2+1}{\Im  \, \Phi}=\xi$, &   $\tilde\Phi =0$  & $2<\xi$\\
 & & & \\
\hline
\end{tabular}
\end{center}
\caption{Type of vacua of the scalar potential for different values of  the parameter  $\xi$. 
Non-singular cosmic string solutions are only possible for  $2<\xi$.\label{Minkowski}
 }
\end{table}

When $\xi\leq 0$ there are no Minkowski vacua. This implies in particular that for a  gauging with  $\xi\leq 0$, all the extrema of the potential are de Sitter vacua. However, the potential will not have an absolute minimum  (for finite values of the fields) because of its runaway behaviour in the dilaton   \eqref{runAway}.\\

We shall use table \ref{Minkowski} to explain our choice for the cosmic string configuration.
In order to have a cosmic string solution we need to have a circle in the vacuum manifold. 
If we want the string configuration to be compatible with a  consistent reduction of supersymmetry, we shall have to truncate some of the scalar fields of the quaternionic manifold to end up with a K\"ahler-Hodge submanifold which is completely geodesic. \\

The  appropriate choice of gauging to construct a cosmic string of the Nielsen-Olesen type is  $\xi>2$. 
 Indeed, in that case, the vacuum is a circle defined by $\frac{\abs{\Phi}^2+1}{\Im  \, \Phi}=\xi$.
   The Higgs field of the cosmic string is $\Phi$.  We shall keep $y=\tilde\Phi=0$ not only in the vacuum but for all the string solutions in order to  have a consistent truncation.    \\

In the case $0<\xi<2$, we also have a circle in the vacuum. However, $\Phi=\frac{\xi}{2}\ii$ does not define a consistent truncation of the quaternionic manifold. To see this note that a gauge transformation
\ref{gaugetransf} with the killing vector $k$ given by (\ref{killing})  does not respect this condition for every value of $\tilde \Phi$.\\
In the case where $\xi=2$, the vacuum is just a point and therefore there is no room for a cosmic string solution of the Nielsen-Olesen type.

\subsection{Consistent truncation}

The set of conditions that we impose 
on the bosonic fields defining the consistent reduction are:
\begin{equation}
\text{Consistent reduction ansatz : } \quad 
\begin{cases}
y &= 0, \\
\tilde\Phi & =0, \\
W_0=W_1 &=0. 
\end{cases}
\label{truncation}
\end{equation}

The condition $y=\tilde\Phi=0$ was explained in the previous section. 
The conditions $W_0=W_1=0$ ensure that the graviphoton   (see equation \eqref{Graviphoton}) vanishes as it should be in a consistent truncation to $\cN=1$ supergravity. Indeed, the graviphoton appears in the  supersymmetry transformations of the gravitini in $\cN=2$ supergravity  \eqref{susytransformations} but is absent in those of  the gravitino of $\cN=1$ supergravity \eqref{N=1Susy}. \\

In this field configuration the quaternionic and special K\"ahler  manifold reduce 
as follow: 
\begin{equation}
\cM_{SK}\times\cM_{Q} \overset{y=\tilde \Phi=0}{\Longrightarrow} \left({\frac{\SU(1,1)}{\U(1)}}\right)_S\times \left({\frac{\SU(1,1)}{\U(1)}}\right)_\Phi \simeq 
\left({\frac{\SO(2,2)}{\SO(2)\times \SO(2)}}\right)_{S,\Phi},
\end{equation}
with K\"ahler potential 
\begin{equation}
\cK= -\log\left[{-\ii(S-\bar S)}\right] -2\log \left[{-\ii (\Phi -\bar \Phi)}\right].
\end{equation}
Here $S$ is an axion-dilaton field and $\Phi$ is the scalar field whose Higgs mechanism generates  the cosmic string. 
Once we impose the condition $\tilde\Phi=0$, the Killing vector of the quaternionic manifold that we have gauged  only acts on $\Phi$ as:
\begin{equation}
\delta \Phi = - 2g(\Phi^2+1),
\end{equation}
which corresponds to the compact $\U(1)$ symmetry of $\left({\frac{\SU(1,1)}{\U(1)}}\right)_\Phi $.

\subsection{Truncated $\cN=1$ Lagrangian and supersymmetry transformations.}

The bosonic sector of the $\cN=2$ supergravity action is: 
\begin{eqnarray}
e^{-1}\cL&=& \frac{1}{2} R 
 + \frac{1}{4} (\Im \, \cN)_{IJ} F^{I \mu\nu}F^J_{\mu\nu} 
+\frac{e^{-1}}{8}(\Re \, \cN)_{IJ} \varepsilon^{\mu\nu\rho\sigma}F^I_{\mu\nu}F^J_{\rho\sigma}\nonumber\\
& & 
-\frac{1}
{(2 \Im S)^2} \partial _\mu S \partial  ^\mu \bar S-\frac{2}{(1-y \bar y)^2}\partial _\mu y \partial  ^\mu \bar y  
\nonumber \\
&& {}  -\ft12 g_{XY}\nabla _\mu q^X\nabla ^\mu q^Y \nonumber \\
&&+ \frac{2g^2}{\Im S}\left[ 2 k^2 \frac{y \bar y}{(1-y \bar y)^2} +\cP^x \cP^x    \right]^2,
\end{eqnarray}
where:
\begin{equation}
 (\Im \, \cN)_{IJ}= \Im \, S\,  \unity_{3} \qquad (\Re \, \cN)_{IJ}=  \begin{pmatrix}
\Re \, S  \unity_{2} & \\ 
& -\Re \, S,
\end{pmatrix}
\end{equation}
and the metric $g_{XY}$ is given by (\ref{hypermetric}). The hyperscalars are organized as $q^X=(h,\vec b)$. The covariant derivatives are defined
by (\ref{covder}) and the killing vector (\ref{killing}) is gauged by $W^2$. The square of  the moment map, $\cP^x \cP^x$ is given in (\ref{momentmap2}).\\

After setting to zero the truncated fields (\ref{truncation}) and imposing $\Re(S)=0$  the bosonic sector of the $\cN=1$ reduced action reads:
\begin{eqnarray}
e^{-1}\cL&=& \frac{1}{2} R  +\frac{\Im S}{4}F^{\mu\nu}F_{\mu\nu} -\frac{1}
+\frac{e^{-1}}{8}\Re S \varepsilon^{\mu\nu\rho\sigma}F_{\mu\nu}F_{\rho\sigma}\nonumber\\
& &-\frac{1}
{(2 \Im S)^2} \partial _\mu S \partial  ^\mu \bar S-\frac{1}
{2(\Im \Phi)^2} \nabla  _\mu \Phi\nabla  ^\mu \bar \Phi \nonumber \\
&& -2  \frac{g^2}{\Im S}\left[       \frac{\abs{\Phi}^2+1}{\Im \Phi}-\xi  \right]^2,
\end{eqnarray}
where
\begin{equation}
  \nabla _\mu\Phi =\partial _\mu \Phi-2gW_\mu \left( \Phi ^2+1\right).
 \label{nablaPhi}
\end{equation}
In the truncated theory the following relations hold:
\begin{subequations}
\begin{align}
S^{ij} &= T^-_{\mu\nu} = N_{ij}^S=\cN^{i A}=0,\\
\rme^{\cK} &= \frac{1}{-\Im S},\\ 
\cD_S X^I &= \frac{\rme^{\frac{3}{2}\cK}} {4}
\begin{pmatrix}-\ii\\ 1\\0\end{pmatrix}
, \quad \cD_{y^a} X^I =\rme^{\cK/2}\delta^I_a
,
\\
V_i{}^j &= \ii (\omega^3+g W \cP^3)_i {}^j,  \\
  N_{ij}^{y}&=- \rme^{\cK/2}\cP^3_{ij},\\
A & =  -\frac{\ii}{4}\frac{\rmd S+\rmd \bar S}{\Im S}.
\end{align}
\end{subequations}

The previous relations  are useful to compute the supersymmetry  transformations of the fermions from equation \eqref{susytransformations}:

\begin{align}
\delta \psi_{\mu}^1 &=  (\partial_\mu +\ft{1}{4}\omega_
{\mu|ab}\gamma^{ab}+ \ft12\ii A_\mu+\ft12\ii A^B_\mu )\,\epsilon^1,  & \delta
\psi_{\mu}^2 &=  (\partial_\mu +\ft{1}{4}\omega_
{\mu|ab}\gamma^{ab}+\ft12\ii A_\mu-\ft12\ii A^B_\mu )\,\epsilon^2,\nonumber
\\
 \delta\lambda_2^y & =
- \frac{\rme^{\frac{\cK}{2}}}{2}(\Im S) F_{12}\gamma^{12}\epsilon^1
-\ii g\, \rme^{\frac{\cK}{2}} \cP^3\, \epsilon^1,
  &
\delta \lambda_1^y & =  
 \frac{\rme^{\frac{\cK}{2}}}{2}(\Im S) F_{12}\gamma^{12}\epsilon^2
-\ii g\, \rme^{\frac{\cK}{2}} \cP^3\, \epsilon^2,
\nonumber
\\
\delta \lambda^S_1 &=\slashed{\nabla} S \epsilon_1, &  \delta \lambda^S_2 &=\slashed{\nabla} S \epsilon_2, \nonumber\\
 \delta \zeta^1 &=  \frac{\sqrt{2} }{4\Im\, \Phi}\slashed{\nabla} \Phi
\epsilon_1,   &
 \delta \zeta^2 &= - \frac{\sqrt{2} }{4\Im\, \Phi}\slashed{\nabla}\bar \Phi
\epsilon_2. 
\label{susytrans}
\end{align}
In these equations  $A^B_\mu$ is the  quaternionic matter connection of the gravitini:
\begin{equation}
A^B_\mu=2\omega^3_\mu +2 g W_\mu \cP^3=\frac{2 ^2\left( \partial _\mu \Phi
+\partial _\mu\bar \Phi\right)  }{4\Im\Phi }+2g W_\mu \cP^3.
 \end{equation}
and $A_\mu$ is the $\U(1)$ connection of the special K\"ahler manifold.

The main difference with  the supersymmetry transformations obtained in \cite{N2DtermString} on the string configuration  is the presence of the axion-dilaton field $S$ coming from the special geometry and parametrizing the manifold $\frac{\SU(1,1)}{\U(1)}$. 
The gaugini  $\lambda^S_i$ and  the $\U(1)$ connection $A$ of the axion-dilaton scalar manifold do not distinguish between the two supersymmetry transformations :
\begin{itemize}
\item  In the  supersymmetry transformations of the gravitini fields,  the $\U(1)$ connection  $A$  appears with the same charge for both transformations whereas  the matter connection $A^B$ (coming from the $\SU(2)$ of  the hypermultiplet)  comes with opposite charge for the  supersymmetries. 
\item The axion-dilaton  field $S$ enters in the same way in the supersymmetric transformations  of the gaugini $\lambda^S_i$ in contrast to the way $\Phi$  appears in the supersymmetric transformation of the hyperini. 
\end{itemize}

This difference of behaviour will be more clear in the next section where we analyze the different BPS projectors obtained from the BPS equations.

\subsection{Profile of the string}
The BPS equations are obtained by setting to zero the supersymmetry transformations (\ref{susytrans}).\\
In  appendix A, we show that a half-BPS solution for a cosmic string solution with magnetic flux requires that $\epsilon^1$ and $\epsilon^2$ should have different chirality on the cosmic string world sheet:
\begin{equation}
\text{BPS projector:}\quad \gamma^{12} \epsilon^1=\mp \ii\epsilon^2, \quad \gamma^{12} \epsilon^2 =\pm\ii \epsilon^2.
\end{equation} 
The  integrability condition  is
\begin{equation}
 R_{\mu\nu\, 12}\pm  F^B_{\mu\nu} =0.
\end{equation}
Using the  previous projector the BPS equations read
\begin{align}\label{BPSEquations}
(\partial_\mu \mp \frac{\ii}{2}\omega_{\mu|12} +\frac{\ii}{2} A^B_\mu) \epsilon^1 & =0 ,\nonumber  \\
(\partial_\mu \pm \frac{\ii}{2}\omega_{\mu|12} -\frac{\ii}{2} A^B_\mu) \epsilon^2 & =0 ,\nonumber \\
\mp F_{12}+ D &=0 , \nonumber  \\
(\nabla_1 \pm \ii \nabla _2 )\Phi &=0,
\end{align}
where $\quad D= 2\frac{g}{\Im S} \cP^3=-2g\rme^{-\rho} \cP^3$. In  appendix A, it is also shown that the BPS equation for the axion-dilaton field implies that it has to be a constant.\\

The BPS equations \eqref{BPSEquations} are the same as those  obtained in  \cite{N2DtermString} modulo the factor of $\rme^{-\rho}$ in the definition of the $D$-term. \\

Since the coupling of the Higgs field to the gauge field is non standard (\ref{covder}), it is difficult to 
see what would be the field configuration that corresponds to a cosmic string.
To simplify the analysis we define, as in  \cite{N2DtermString}, the following field:
\begin{equation}
u=\frac{i-\Phi}{i+\Phi}.
\end{equation}
Under a gauge transformation the field $u$ transforms as:
\begin{equation}
\delta u = 2 g i u,
\end{equation}
 which corresponds to a change of phase. Thus the winding of the phase of $u$ is the one inducing the magnetic flux of the string.\\

In order to solve the BPS equations we will use the following time independent ansatz:
\begin{equation}
u= f(r) e^{i m \theta} \qquad W_\theta = W_\theta(r).
\end{equation}
It represents a straight cosmic string of winding $m$ along the $z$-axis.

Where we have used cylindrical coordinates $(t,z,r,\theta )$. \\ 
We take the  space-time metric to be of the form:
\begin{equation}
  \rmd s^2 = -\rmd
t^2+\rmd z^2+\rmd r^2+C^2(r)\rmd\theta^2.
\end{equation}
The BPS equations for the profile of the string are :
\begin{equation}
f'(r) = \pm \frac{f(r)}{C(r)}\left({m-4gW_{\theta}(r)}\right), \qquad
W'_{\theta}(r)=\pm C(r)D(r), \qquad
 C'(r)=1\mp \tilde A^B_{\theta}(r),
\end{equation}
with
\begin{equation}
\tilde A^B_{\theta} = 2 \, m  \frac{f^2}{1-f^2}+ \rme^{\rho}W_{\theta} D \quad \text{and} \quad 
D=-2g \rme^{-\rho}(2 \frac{1+f^2}{1-f^2}-\xi).
\end{equation}

From the BPS equations we find the asymptotic behavior of the profile functions.
It is similar to the cases of \cite{Dvali:2003zh, N2DtermString}.\\
In the case $r \to 0$ we have:
\begin{equation}
f(r) \sim r^{\pm n}, \qquad C \sim r, \qquad W_{\theta}(r) \sim \pm g(\xi - 2) r^2.
\end{equation}
In the opposite limit, $r \to \infty$:
\begin{equation}
f(r) \Longrightarrow \sqrt{\frac{\xi-2}{\xi+2}}, \qquad W_{\theta}(r) \Longrightarrow \frac {m}{4g},
\end{equation} 
and the metric is given by
\begin{equation}
\rmd s^2 = -\rmd t^2+\rmd z^2+\rmd r^2+r^2\left[1\mp \ft12m(\xi-2
)\right]^2\rmd\theta^2.
\end{equation}

The upper or lower sign apply for positive or negative winding number
$m$, respectively.

At $r\Longrightarrow \infty$, the string creates a locally-flat conical
metric with a deficit angle proportional to $\xi -2$.  The energy
of the string per unit length can be computed as in \cite{Dvali:2003zh, N2DtermString}.
One finds that  the only non-vanishing contribution comes from the
Gibbons-Hawking surface term \cite{Gibbons:1976ue}:
\begin{equation}
\mu_{\mbox{string}} = -\left.\int \rmd\theta\, C'\right|_{r=\infty } +
\left.\int \rmd\theta\, C'\right|_{r=0} =\pm  \pi m(\xi-2)>0.
\label{tension}
\end{equation}
Note also that the full $N=2$ supersymmetry is restored asymptotically.

\subsection{The fate of the axion-dilaton field }

The constant value of the axion-dilaton field is not fixed by the BPS equations nor by the scalar potential. The mass per unit length of the string is also independent of the value of the axion-dilaton field (\ref{tension}).
The dilaton fixes the  overall length scale  of the configuration in the following sense.   
There are two natural  lengths in the solution given by the inverse of the masses of the Higgs  and  the gauge field, and they are both functions of the dilaton field: 
\begin{equation}
m_{W}^2\propto -\frac{1}{\Im S}, \quad m_\Phi^2\propto -\frac{1}{\Im S},
\end{equation}
so that the corresponding length scales are 
\begin{equation}
l_{W}^2\propto -\Im S, \quad l_\Phi^2\propto -\Im S.
\end{equation}

Suppose we have a solution to the BPS equations given by the profile functions $f(r)$, $W_\theta(r)$, $C(r)$ and 
$\rho$. Then is easy to check that the functions $f(\lambda r)$, $W_\theta( \lambda r)$, $C(\lambda r)/\lambda$ and  $\rho - 2 \log( \lambda )$ also satisfy the BPS equations for any real $\lambda>0$.
From here is obvious that the value of the dilaton determines the length scales in the transverse direction to the string, in particular the core radius.\\

This situation looks similar to the case of semilocal strings \cite{VachaspatiDZ}, were there is also a one parameter family of solutions with equal energy and different core radii. In that case finite energy perturbations can 
excite the zero mode connecting solutions within the same family, leading to the spread of the magnetic flux and eventually to the disappearance of the strings.
This is not going to occur in our model.
In order to go from one solution to a different one, the dilaton has to change its value everywhere in the plane transverse to the string. The kinetic energy needed in order to excite the value of the dilaton globally diverges, and this implies that, once the system has chosen a given value for the dilaton,  finite energy perturbations cannot drive the system to a solution with a different value of $S$. The radius of the string will remain unchanged.

\section{Discussion}

As a generalization of the work done in \cite{N2DtermString},
we have enlarged the family of $\cN=2$ supergravity actions which allow the embedding of $\cN=1$ supergravity actions containing a $D$-term potential and a constant FI term.
We extend the result of \cite{N2DtermString} to a class of special geometries more familiar  in compactifications of string theory. We are using here a ``very special K\"alher geometry'' characterized by a cubic prepotential, instead of the minimal special geometry used in \cite{N2DtermString}. To be specific we take the special manifold to be:
\begin{equation}
ST[2,n] \equiv \frac{\SU(1,1)}{\U(1)}\times \frac{\SO(2,n)}{\SO(2)\times \SO(n)}, \nonumber
\end{equation} 
in the Calabi-Visentini basis \eqref{CV}, which is related to the cubic prepotential by a symplectic rotation \cite{Andrianopoli:1997cm,Ceresole:1995jg}.\\

This choice of special geometry has two important consequences.
An axion-dilaton field, $S=a-\ii\rme^\rho$,  is present in the reduced $\cN=1$ theory after truncation from $\cN=2$. Moreover,   it is possible to define a gauging for which the scalar potential is bounded from below.   However, it  has a runaway dependence on the dilaton: 
\begin{equation}
\cV\propto \rme^{-\rho}.\nonumber
\label{Dpotential}
\end{equation}

As an application, we have shown how to construct a half-BPS cosmic string solution from a $\cN=2$ supergravity action in $D=4$. Following  \cite{N2DtermString} we have used a string ansatz compatible with a consistent truncation from $\cN=2$ to $\cN=1$. In order to obtain the scalar potential we have gauged  the same isometry used in  \cite{N2DtermString}. 
We have found that the BPS equations imply that the axion-dilaton has to be simultaneously holomorphic and anti-holomorphic, which can only be satisfied if it is a constant:
\begin{equation}
S= \text{Constant}, \quad {\rm Im} S <0. \nonumber
\end{equation}  
Despite the runaway behavior of the potential, we have proved that all the string solutions have the same energy per unit length, regardless of the value of the dilaton, and it is given by  the Gibbons-Hawking surface term \cite{Comtet:1988wi,  Craps:1997gp, N2DtermString}. The value of the dilaton fixes the masses  of the Higgs and the gauge field and, hence, also the radius of the string. 
We have argued that the system can not evolve between two solutions with different values of the dilaton, since this would require an infinite amount of energy. Thus, once the strings are formed their radii remain fixed.\\

Observations of the timming of milisecond pulsars give the constraint  $\mu_{\mbox{string}} \lesssim 10^{-7}$  \cite{Davis:2005dd}. However this constraint depends on the specific model used to calculate it, what leads to a considerable uncertainty. 
This implies for our model that the FI term has to satisfy:
\begin{equation}
0<2 \pi m ( \xi-2) \lesssim10^{-7},\nonumber
\end{equation}
where the lower bound is coming from the study of Minkowski vacua in section 5.1.\\

An important issue which remains to be discussed is the stability of these strings within the full $\cN=2$ theory. It is far from clear that these strings are stable against perturbations of the truncated fields.
Finding such an instability would be an interesting result, since it would provide an example of an unstable BPS solution.

\medskip
\section*{Acknowledgments.}

\noindent We are grateful to L. Leblond, J. Polchinski,  M. Ro\v{c}ek,  S.E. Shandera,  S. Vandoren, B. de Wit  and specially  to A. Ach{\'u}carro, S. Davis, and  R. Jeannerot  for very useful  discussions. M.E. would like to express his gratitude to the participants and organizers of the 4th Simons Workshop in Mathematics and Physics at Stony Brook for their hospitality and numerous discussions. M.E.  also  thanks  the 2006  Prospects in Theoretical Physics  for the wonderful school on ``Applications of String Theory"  at the Institute for Advanced Studies in Princeton where part of this work was done.   
This work is partly supported by the Basque Government  grant BFI04.203.
\\
\medskip

\appendix
\section{Integrability condition and BPS projector.}

The integrability conditions of the gravitini BPS equations are\footnote{The integrability conditions of the gravitini  are obtained as usual by taking the commutator of two supersymmetries and using the relation 
$$
[\nabla_\mu , \nabla_\nu]= R_{\mu\nu} .$$ 
  We shall name $F^\cK$ the curvature  of the special geometry $\U(1)$ connection $A$ and $F^B$ the curvature of the connection $A^B$. }:
\begin{equation}
(R_{\mu\nu\, 12}\gamma^{12}+\ii F^{\cK}_{\mu\nu}+\ii F^B_{\mu\nu} )\epsilon^1=0, \quad ( R_{\mu\nu\, 12}\gamma^{12}+\ii F^{\cK}_{\mu\nu}-\ii F^B_{\mu\nu} )\epsilon^2=0,
\end{equation}

To compute the projector we use the following result\footnote{Given two complex variables $a$ and $b$ and a spinor $\epsilon$, we have 
 $$(a\gamma^1\unity +b\gamma^2)\epsilon=0\Longrightarrow (a\gamma^1 +b\gamma^2)^2\epsilon=0\Longrightarrow (a^2+b^2)\epsilon=0 \overset{\epsilon \neq 0}{\longrightarrow} a\mp \ii b=0.
$$ Thus: 
$$
(a\gamma^1 +b\gamma^2)\epsilon=0 \overset{a\neq 0}{\Longrightarrow} (\gamma^1\mp \ii \gamma^2)\epsilon =0,
$$}:
\begin{equation}
\left.{
\begin{matrix}
(a+\gamma^{12}b)\epsilon=0\\
\text{ or }\\
(a\gamma^1 +b\gamma^2)\epsilon=0
\end{matrix}
}\right\}
 \overset{\epsilon \neq 0, a \neq 0}{\Longrightarrow} a \mp \ii b=0 , \quad \gamma^{12}\epsilon= \mp \ii \epsilon.
\end{equation}

Using the previous  method we obtain
\begin{subequations}
\begin{align}
\begin{split}
\mp R_{\mu\nu\, 12}+ (F^{\cK}_{\mu\nu}+F^B_{\mu\nu}) &=0, \quad \gamma^{12} \epsilon^1 = \mp\ii  \epsilon^1,    \\
\mp R_{\mu\nu\, 12}+ (F^{\cK}_{\mu\nu}-F^B_{\mu\nu}) &=0,  \quad \gamma^{12} \epsilon^2 = \mp\ii  \epsilon^2, \label{GravitiniProj}
\end{split}
\end{align}
\begin{align}
\begin{split}
\mp F_{12}+ D &=0 ,\quad \gamma^{12} \epsilon^1 = \mp\ii \epsilon^1,    \\
\mp F_{12}- D &=0 ,\quad \gamma^{12} \epsilon^2 = \mp\ii  \epsilon^2, \label{PhiGauginiProj}
\end{split}
\end{align}
\begin{align}
(\partial_1 \mp \ii \partial_2 ) S &=0, \quad \gamma^{12}\epsilon^1 =\mp \ii \epsilon^1, \quad \gamma^{12}\epsilon^2 = \mp \ii \epsilon^2,   \label{SGauginiProj}
\end{align}
\begin{align}
\begin{split}
(\nabla_1 \mp \ii \nabla _2 )\Phi &=0, \quad \gamma^{12}\epsilon_1 =\mp\ii  \epsilon_1,    \\
(\nabla_1 \mp \ii \nabla _2 )\bar \Phi &=0, \quad  \gamma^{12}\epsilon_2 = \mp \ii \epsilon_2 ,\label{PhiProj}
\end{split}
\end{align}
\end{subequations}
where $ D= 2\frac{g}{\Im S} \cP^3=-2g \rme^{-\rho} \cP^3$.

Equations  \eqref{GravitiniProj} are the integrability conditions for the gravitini BPS equations. 
They imply that  $F^{\cK}$ (resp.  $F^{B}$ ) vanishes if  $\epsilon^1$ and $\epsilon^2$ have opposite (resp. the   same ) chirality on the string world sheet: 
\begin{subequations}
\begin{align}
\begin{split}
\gamma^{12}\epsilon^1 &= \mp\ii \epsilon^1 \text{ and }\gamma^{12}\epsilon^2= \mp \ii\epsilon^2 \Longrightarrow  F^{B}=0, \\
\gamma^{12}\epsilon^1 &= \mp\ii \epsilon^1 \text{ and }\gamma^{12}\epsilon^2= \pm \ii\epsilon^2 \Longrightarrow 
F^{\cK}=0, 
\end{split}
\end{align}
\end{subequations}
Equations  \eqref{PhiGauginiProj} are  compatible in a non-trivial situation (that is for $F_{12}\neq 0$ and $D\neq 0$) if and only if $\epsilon^1$ and $\epsilon^2$ have different chirality on the string world sheet as: 

\begin{align}
\begin{split}
\gamma^{12}\epsilon^1 &= \mp \ii\epsilon^1 \text{ and }\gamma^{12}\epsilon^2= \mp \ii\epsilon^2 \Longrightarrow 
F_{12}=D=0, \\
\gamma^{12}\epsilon^1 &= \mp \ii\epsilon^1 \text{ and }\gamma^{12}\epsilon^2= \pm \ii\epsilon^2 \Longrightarrow 
\mp F_{12}-D=0.
\end{split}
\end{align}
Equations \eqref{SGauginiProj} implies that $S$ is a constant if  $\epsilon^1$ and $\epsilon^2$ have different chirality whereas $S$  is holomorphic or anti-holomorphic  on the plane perpendicular to the axis of the string when $\epsilon^1$ and $\epsilon^2$   have the same chirality on the world sheet: 
\begin{align}
\begin{split}
\gamma^{12}\epsilon^1 &= \mp \ii\epsilon^1 \text{ and }\gamma^{12}\epsilon^2= \mp\ii \epsilon^2 \Longrightarrow 
  (\partial_1\mp \partial_2)S=0, \\
\gamma^{12}\epsilon^1 &= \mp\ii \epsilon^1 \text{ and }\gamma^{12}\epsilon^2= \pm\ii \epsilon^2 \Longrightarrow  S=\text{Constant},
\end{split}
\end{align}

Finally the hyperini BPS equations are related by complex conjugation. 
It follows that if $\epsilon_1$ and $\epsilon_2$ have different chirality, $\Phi$ is covariantly holomorphic on the plane perpendicular to the string axis. If they have the same chirality it is at the same time holomorphic and anti-holomorphic and therefore constant :
 \begin{align}
\begin{split}
\gamma^{12}\epsilon^1 &= \mp\ii \epsilon^1 \text{ and }\gamma^{12}\epsilon^2= \pm\ii \epsilon^2 \Longrightarrow   (\nabla_1\mp\ii \nabla_2)\Phi= 0,\\
\gamma^{12}\epsilon^1 &= \mp \ii\epsilon^1 \text{ and }\gamma^{12}\epsilon^2= \mp\ii \epsilon^2 \Longrightarrow 
  \Phi=\text{ Constant}.
\end{split}
\end{align}

As we are interesting in a cosmic string solution involving a magnetic flux ($F\neq 0$) and a non-trivial scalar field $\Phi$ ($\neq \text{Constant}$) we shall take  the projector which ensures that $\epsilon^1$ and $\epsilon^2$ have different chirality on the string world sheet:
\begin{equation}\label{BPSProj}
\text{BPS projector:}\quad \gamma^{12} \epsilon^1=\mp \ii\epsilon^2, \quad \gamma^{12} \epsilon^2 =\pm\ii \epsilon^2.
\end{equation}

If follows that the axion-dilaton is a constant on the string configuration and that the $\U(1)$ connection  ($A$) of the special geometry and its curvature $F^{\cK}$  both vanish identically :
\begin{equation}
 A=F^{\cK}=0.
\end{equation}


\end{document}